\documentclass[10pt,journal]{IEEEtran}
\usepackage[utf8]{inputenc}
\usepackage{url}
\usepackage{verbatim}
\usepackage{enumitem}
\usepackage{amsmath}
\usepackage{mathtools}
\usepackage{listings}
\usepackage{color}
\usepackage{subcaption}
\usepackage{cite}
\usepackage{lipsum}
\usepackage{graphicx}
\usepackage{mathcomp}
\usepackage{hyperref}
\usepackage[british]{babel}
\usepackage{expl3}
\usepackage{environ}
\usepackage{textcomp}
\usepackage{boxhandler}

\usepackage{etoolbox}

\let\oldtextbf\textbf
\renewcommand{\textbf}[1]{\oldtextbf{\boldmath #1}}

\definecolor{mygreen}{rgb}{0,0.6,0}
\definecolor{mygray}{rgb}{0.5,0.5,0.5}
\definecolor{mymauve}{rgb}{0.58,0,0.82}

\lstdefinestyle{example}{
  float=tp,
  floatplacement=tbp,
  abovecaptionskip=-5pt,
  numberstyle=\fontsize{7}{9}\selectfont\ttfamily\bfseries,
  numbers=left,
  numbersep=8pt, 
  xleftmargin=2em,
  frame=tb,
  framexleftmargin=1.5em,
  language=C, 
  basicstyle=\fontsize{7}{9}\selectfont\ttfamily,
  breaklines=true
}

\newcommand{\hypgeo}[2]{%
  \operatorname{%
    {\vphantom{\mathnormal{F}}}_{#1}%
    \kern-\scriptspace
    \mathnormal{F}_{#2}%
  }%
}
\usepackage[T1]{fontenc}
\usepackage{babel}
\begin{document}
\title{Metronome: adaptive and precise intermittent packet retrieval in DPDK}




\author{Marco Faltelli$^{1}$, Giacomo Belocchi$^{1}$, Francesco Quaglia$^{1,2}$, Salvatore Pontarelli$^{2,3}$, Giuseppe Bianchi$^{1,2}$
\\
$^1$University of Rome Tor Vergata, Italy.\\
$^2$Consorzio Nazionale Interuniversitario per le Telecomunicazioni (CNIT), Italy.\\
$^3$Sapienza University of Rome, Italy.\\}
\maketitle

\begin{abstract}
The increasing performance requirements of modern applications place a significant burden on software-based packet processing. Most of today's software input/output accelerations achieve high performance at the expense of reserving CPU resources dedicated to continuously poll the Network Interface Card. This is specifically the case with DPDK (Data Plane Development Kit), probably the most widely used framework for software-based packet processing today. The approach presented in this paper, descriptively called Metronome, has the dual goals of providing CPU utilization proportional to the load, and allowing flexible sharing of CPU resources between I/O tasks and applications. Metronome replaces DPDK's continuous polling with an intermittent sleep\&wake mode, and revolves around a new multi-threaded operation, which improves service continuity. Since the proposed operation trades CPU usage with buffering delay, we propose an analytical model devised to dynamically adapt the sleep\&wake parameters to the actual traffic load, meanwhile providing a target average latency. Our experimental results show a significant reduction of the CPU cycles, improvements in power usage, and robustness to CPU sharing even when challenged with CPU-intensive applications.


\end{abstract}

\section{Introduction}
Packet processing is a very common task in every modern computer network, and Data Centers allocate relevant amounts of resources to accomplish it. Also,
DPDK is the most used framework for software packet processing, as it provides excellent performance levels 
\cite{gallenmuller2015comparison}. On the downside, deploying DPDK applications comes with a series of shortcomings, the major one being the need for fully reserving  a subset of the available CPU-cores for continuously polling the NICs---a choice that has been made in order to timely process incoming packets. This approach not only gives rise to constant, 100\% utilization of the reserved CPU-cores, but also leads to high power consumption, regardless of the actual volume of packets to be processed \cite{xu2016demystifying}.

Indeed, there are many reasons which suggest that the availability of solutions capable to replace continuous polling with an intermittent, sleep\&wake, CPU-friendly approach would be beneficial. While Google states that even small improvements in resources utilization can save millions of dollars \cite{borg}, previous work has brought about evidence that despite Data Center networks are designed to handle peak loads, they are largely underutilized. Microsoft reveals that 46-99\% of their rack pairs exchange no trafﬁc at all \cite{ghobadi2016projector}; at Facebook the utilization of the 5\% busiest links ranges from 23\% to 46\% \cite{roy2015inside}, and \cite{benson2010network} shows that the percentage of utilization of core network links (by far the most stressed ones) never exceeds 25\%. Fully dedicating a CPU, the most greedy component in terms of power consumption \cite{niccolini2012building, datacenterbook}, to continuous NIC polling thus appears to be a significant waste of precious resources that could be exploited by other tasks. To a greater extent this appears the case nowadays: CPU performance is struggling to improve and seems about to reach a stagnation point \cite{hennessy2019new, moorezil-reconf-net-sys}, at the moment of time in which CPUs burden is ever growing, also because of newly emerging needs for security  (e.g. the Kernel Page Table Isolation---KPTI--facility adopted by Linux to prevent attacks based on hardware level speculation, like Meltdown \cite{meltdown}). 

DPDK's continuous CPU usage may also raise concerns in multi-tenant cloud-based deployments, where customers  rent virtual CPUs which are then mapped onto physical CPUs in a time-sharing fashion. In fact, fully reserving CPUs for DPDK tasks complicates (or makes unfeasible) the adoption of resource sharing between different cloud customers. 
Also, 100\% usage of computing units is not favorable to performance in scenarios where threads run on hyper-threaded machines---just because of conflicting usage of CPU internal circuitry by the hyper-threads. Hence, multi-threading should be avoided in continuous polling-based DPDK deploys, posing the additional problem of making this framework not fully prone to scale on off-the-shelf parallel machines. While major cloud providers \cite{ENA, azure} have already enabled the deployment of DPDK applications in their data centers, to the best of our knowledge such solutions still present the shortcomings of drivers based on continuous-poll operations.

To face these issues, this paper proposes Metronome \cite{metronome}, an approach devised to replace the continuous DPDK polling with a sleep\&wake intermittent approach. Albeit this might seem in principle an obvious idea, its  advantages are linked to several factors that we cope with in this article.
First, a suited implementation/usage of sleep\&wait operating system services needs to be put in place. As for this aspect, Metronome can work effectively by relying on microsecond level sleep phases 
supported by either the Linux \texttt{nanosleep()} service or our own new service called \texttt{hr\_sleep()} \cite{metronome}. The latter offers a few advantages and 
is also independent of limitations related to system parameterization and thread priorities.
Second, Metronome revolves around a novel architecture and operating mode for DPDK, where incoming traffic, from either a single receive queue or multiple ones, is shared between multiple threads---as we will discuss this also offers advantages by the side of robustness versus operating system thread-scheduling decisions. These threads dynamically switch---in a coordinated manner---from polling the receive queue to sleep phases for short and tunable periods of time when the queue is idle. Owing to a suitable adaptation strategy which tunes the sleeping times depending on the load conditions, Metronome achieves a stable tunable latency and no substantial packet loss difference compared to standard DPDK while reaching significant reduction for both CPU usage and power consumption. 

Overall, the contributions we provide in this article can be summarized as follows:
\begin{itemize}

 \item exploiting a fine grain sleep\&wake service---in particular the {\tt hr\_sleep()} service---we define the Metronome multi-threaded architecture for DPDK
 applications, based on extremely low thread-coordination overhead. It boils down CPU usage compared to classical DPDK settings, and offers a better capability to exploit hardware level parallelism. As an indirect effect, Metronome 
 positively impacts  energy  efficiency  under  specific workloads;
    \item we present an analytical model for Metronome, which is used for driving allocation of CPU to threads,
    making the DPDK framework  dynamically adapt its behavior (and its demand for resources) to the workload;
    \item we extensively assess Metronome on 10 Gbps NICs, in various load conditions, and we test its integration in three different applications: L3 forwarding, IPsec, and FloWatcher \cite{flowatcher}, a high-speed software traffic monitor;
    \item we extend  
    the evaluation of Metronome to 40 Gbps NICs, where multiple receive queues need to be used and therefore, orchestrated by the Metronome algorithm.
\end{itemize}
Metronome is publicly available at \cite{repo}.
\section{Related work}

When processing the packet flow incoming from NICs, two orthogonal approaches can be exploited: (continuous) polling and interrupt. 
Polling-based frameworks can either rely on a kernel driver (e.g. netmap \cite{netmap}, PFQ \cite{pfq} and PF\_RING ZC \cite{zerocopy}) or bypass the kernel through a user space driver, like DPDK \cite{dpdk} and Snabb \cite{snabb}. Such frameworks rely on high performance, batch transferring mechanisms such as DMA and zero copy \cite{zerocopy}, preallocating memory through OS hugepages.
Among all of these solutions, DPDK has definitely emerged as the most used one, as it reaches the best performance levels  \cite{gallenmuller2015comparison}. Furthermore, it is continuously maintained by the Linux Foundation and other main contributors (e.g., Intel).

As mentioned, one of the main shortcomings of DPDK is the excessive usage of resources (CPU cycles and energy), caused by the busy-wait approach used by threads to check the state of NICs and Rx queues. Intel tried in \cite{dpdkwhitepaper} to address the energy consumption issue via a gradual decrease of the CPU clock frequency under low traffic for a commonly used application such as the layer-3 forwarder. 
A similar approach is used in \cite{powerefficientio}, with the addition of an analytical model exploited  to choose the appropriate CPU frequency. Along this line, \cite{niccolini2012building} proposes a power proportional software-router.

However, while the downgrading of the clock frequency reduces power consumption \cite{niccolini2012building} without noticeably affecting performance, these solutions do not take into account another crucial aspect, namely the actual usage of CPU. In fact, downgrading the clock frequency of a CPU-core fully dedicated to a thread operating in busy-wait (namely, continuous polling) mode still implies 100\% utilization. Hence, the CPU-core is anyhow unusable for other tasks. 
Moreover, downgrading the clock frequency of CPUs
is not feasible in cloud environments since (i) they are shared between different processes and customers and (ii) providers would like them to be fully utilized in order to reach peak capacity on their servers \cite{firestone2018azure}. Our proposal bypasses these limitations since we do  not rely on any explicit manipulation of the frequency and/or power state of the CPUs. Rather, we 
control at fine grain the timeline of CPU (and energy) usage by DPDK threads---hence the name Metronome---which are no longer required to operate in busy-wait style. Such control is based on an analytical model, that allows taking runtime decisions depending on packet workload variations.

At the opposite side, the literature offers interrupt-based solutions. However, the huge improvements of NICs (1GbE to 100 GbE), and the contextual stall of CPU performance because of the end of Moore's Law and Dennard Scaling \cite{hennessy2019new, moorezil-reconf-net-sys}
has evidenced performance limitations of the interrupt-based approach. In fact, interrupt-based solutions suffer from the latency brought by the system calls used to interact with the kernel level driver managing interrupts, packet copies to user space and so on. Moreover, an interrupt-based architecture operating at extreme interrupt arrival speed may cause livelocks \cite{livelock}. The Linux NAPI aims at tackling these limitations by providing an hybrid approach which tends to eliminate receive livelocks by dynamically switching between polling and interrupt-based packet processing, depending on the current traffic load. Such a mechanism currently works only for kernel-based solutions, not for user space ones, like DPDK.
There is a growing interest in XDP \cite{xdp, hxdp}, a framework built inside the Linux kernel for programmable packet processing. Instead of moving control out of the kernel (with the associated costs), XDP acts before the kernel networking stack takes control so as to achieve latency reduction. While XDP provides some significant benefits such as total integration with the OS kernel, improved security and CPU usage proportional to the actual network load, it still does not match DPDK's performances (\cite{xdp} and Sec.\ref{sec:xdp}) and currently supports less drivers than DPDK does \cite{drivers_xdp,drivers_dpdk}. Our solution is instead fully integrated with DPDK. Works like Shenango \cite{shenango} and ZygOS \cite{zygos} explicitly target latency sensitive applications, while other contributions try to accelerate packet processing by moving computation to modern NICs \cite{loom, tonic, softnic, xtra}. 


\section{Metronome architecture}

\subsection{Fine-Grain Thread Sleep Service}
\label{fgss}


The precision of the thread-sleep interval supported by the operating system, is essential for the construction of any solution where the following two objectives need to be jointly pursued: 1) threads must leave the CPU if there is currently nothing to do (in our case by the side of packet processing); 2) threads must be allowed be CPU rescheduled---gaining again control of the CPU---according to a tightly controlled timeline. Point 2) would allow the definition of an architectural support where we can be confident that threads will be able to be CPU dispatched exactly at (or very close to) the point in time where we would like to re-execute a poll operation on the state of a NIC---to determine whether incoming packets need to be processed. On the other hand, point 1) represents the basis for the construction of a DPDK architecture not based on full pre-reserving of CPUs to process incoming packets.

In current conventional implementations of the Linux kernel, the support for (fine-grain) sleep periods of threads is based on the {\tt nanosleep()} system call.
A few limits of this service are related to its dependency on a slack factor assigned to threads, which is checked when they request to sleep. This factor can be controlled using the {\tt prctl()} system call, putting it to the minimal value of 1. If such setting is not adopted, then for any thread that is not in the real-time CPU-scheduling class we have at least 50 microseconds as the slack imposed by the Linux kernel, which makes the awake of the thread less controllable in terms of precision under fine-grain sleep requests. Furthermore, when entering the kernel level execution of {\tt nanosleep()}, the Thread Control Block (TCB) is checked because of the need to determine the current slack value, which makes the service run a few machine instructions to reconcile the real value to be adopted in the sleep phase with the information kept in the TCB.

\begin{figure}[tb]
    \centering
    \subfloat[][1 $\mu$s]{%
        \includegraphics[width=0.32\linewidth]{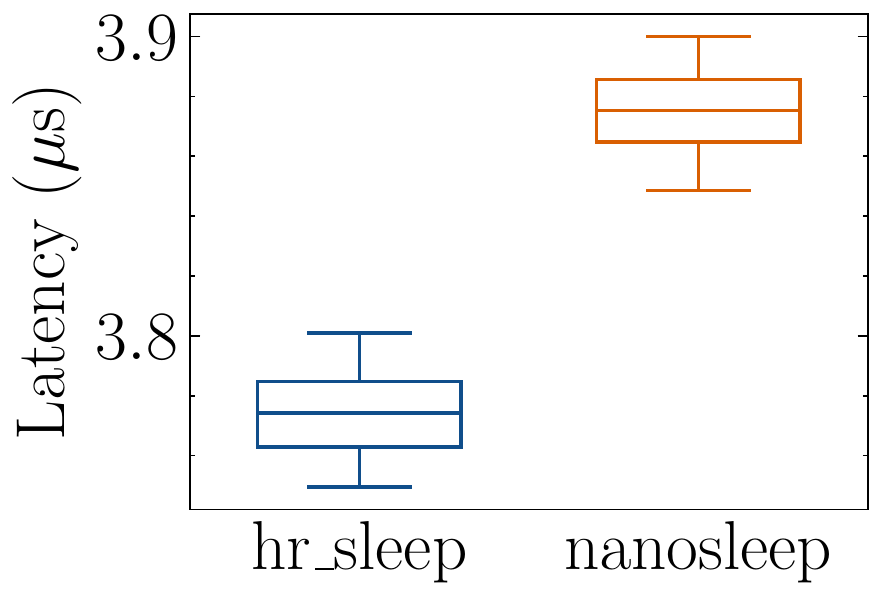}%
        \label{fig:sleep_a}%
        }%
    \subfloat[][10 $\mu$s]{%
        \includegraphics[width=0.35\linewidth]{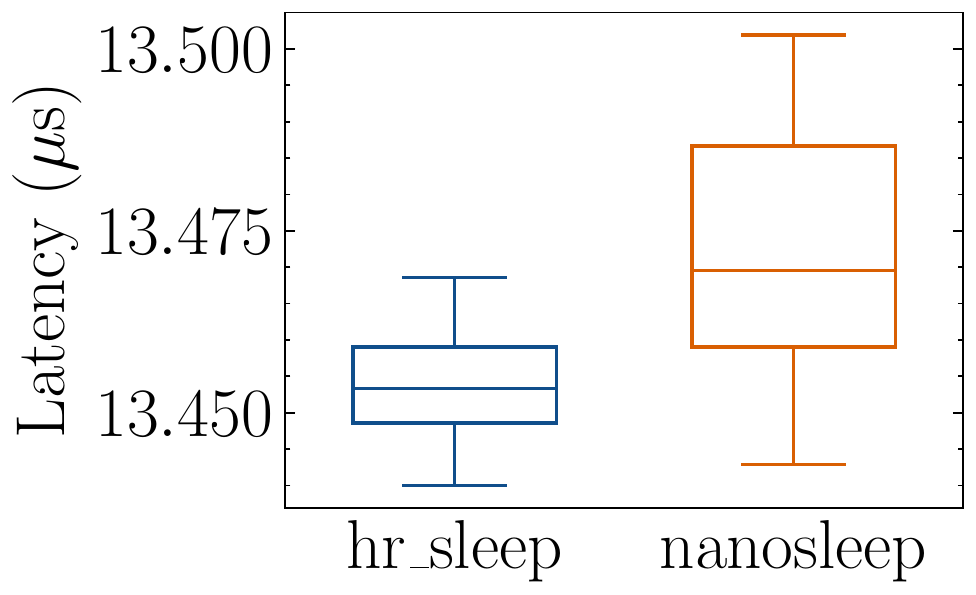}%
        \label{fig:sleep_b}%
        }%
         \subfloat[][100 $\mu$s]{%
        \includegraphics[width=0.34\linewidth]{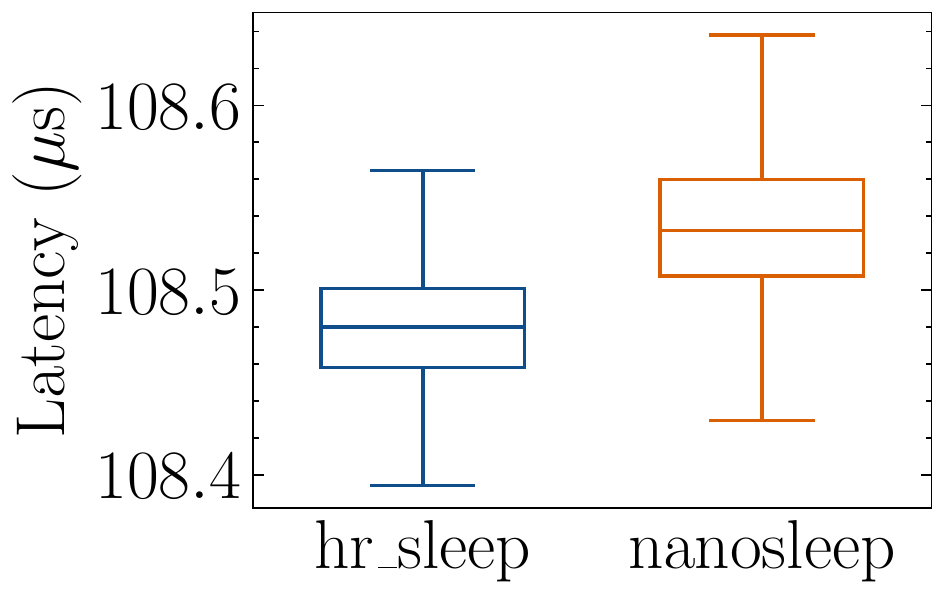}%
        \label{fig:sleep_c}%
        }%
    \caption{Boxplots for \texttt{hr\_sleep()} and \texttt{nanosleep()} latencies} \label{fig:sleep_periods}
\end{figure}

While developing Metronome, we also implemented an alternative sleep service, namely {\tt hr\_sleep()}, whose details are provided in \cite{metronome}. This variant fully avoids any interaction with thread management information at kernel level (such as the current slack value kept in the TCB). Hence, it also avoids running the additional machine instructions needed to manage this information, 
for any CPU-scheduling class of threads (real-time or not). We remand the reader to \cite{metronome} for an extended evaluation of this implementation.
 In any case, 
%
in Figure \ref{fig:sleep_periods} we show the slight advantages 
provided by {\tt hr\_sleep()} even under the scenario where {\tt nanosleep()} is configured with the minimal admissible slack currently supported by the Linux kernel. 
The tests have been conducted on an isolated NUMA node equipped with Intel Xeon Silver 2.1 GHz cores.
The server is running Linux kernel 5.4.
We have run an experiment where a million samples of the wall-clock-time elapsed between the invocation of the sleep-service and the resume from the sleep phase are collected. This wall-clock-time interval has been measured via start/end timer reads operated through \texttt{\_\_rdtscp()}. 
We show the boxplots for both the sleep services with different timer granularity requests (1, 10, 100 $\mu$s). These data have been collected by running the thread issuing the sleep request as a classical {\tt SCHED\_OTHER} (normal) priority thread and---as hinted before---with the timer slack of \texttt{nanosleep()} set to 1$\mu$s.
The results show that \texttt{hr\_sleep()} provides some minimal gains both in terms of mean latency and variance even under such extreme setting of the slack value for {\tt nanosleep()}. In any case, the avoidance of the reliance on kernel level parameters makes {\tt hr\_sleep()} fully independent of any kernel configuration choice for the minimal admissible slack.

\subsection{Actual Thread Operations}

In this section, we describe how threads in charge of processing packets operate in Metronome. To this end, 
let us start with a brief discussion of the state-of-the-art DPDK architecture: on the receiving side, NICs may convey their incoming traffic into a single Rx queue or either split such traffic into multiple Rx queues through RSS. A DPDK thread normally owns (and manages) one or more Rx queues, while an Rx queue belongs to (namely, is managed by) one DPDK thread \cite{multicore_dpdk}. Therefore, the behavior of a DPDK thread is no more than an infinite \texttt{while(1)} loop in which the thread constantly polls all the Rx queues it is in charge of. This approach rises some important shortcomings such as (i) greedy usage of CPU even in light load scenarios (a problem we already pointed to) and (ii) prevention of any Rx queue sharing among multiple threads. As for point (ii) we note that in 40GbE+ NICs, queues experience heavy loads despite the use of the RSS feature,  e.g. on a 100Gb port with 10 queues, each queue can experience 10Gb rate traffic or even more. Preventing multi-threaded operations on each single Rx queue, and the exploitation of hardware parallelism for processing incoming packets from that queue, looks therefore to be another relevant limitation.

Compared to the above described classical thread operations in state of the art DPDK settings, we believe smarter operations can be put in place by sharing a Rx queue among different threads and putting these threads to sleep, when queues are idle, for a tunable period of time, depending on the current traffic characteristics. In other words, via a precise fine-grain sleep service, and lightweight coordination schemes among threads, we can still control and improve the trade-off between resource usage and efficiency of 
packet processing operations.

To this end, the {\tt hr\_sleep()} service
has been coupled in Metronome with a multi-threaded approach to handle the Rx queues. In more detail, in our DPDK architecture we have multiple threads that sleep (for fine grain periods) and then, upon execution resume, race with each other to determine a single winner that will actually take care of polling the  state of some Rx queue for processing its incoming packets. In this approach we do not rely on any additional operating system services to implement the race; rather, we implemented the race resolution protocol purely at user space via atomic Read-Modify-Write instructions, in particular the {\tt CMPXCHG} instruction on {\sf x86} processors, which has been exploited to build a lightweight {\tt trylock()} service. The race winner is the thread that atomically reads and modifies a given memory location (used as the lock associated with an Rx queue), while the others simply 
iterate on calling our new {\tt hr\_sleep()} 
service, thus immediately (and efficiently, given the reduced CPU-cycles usage  of {\tt hr\_sleep()}) leaving the CPU---given that another thread is already taking care of checking with the state of the Rx queue, possibly processing incoming packets\footnote{Interested readers can have a look at Appendix I for a basic coding example of DPDK-traditional and Metronome approaches.}.

We also note that using multiple threads according to this scheme allows creating less correlated awake events and CPU-reschedules, leading to (i) more predictability in terms of the maximum delay we may experience before some Rx is checked again for incoming packets and (ii) less work to be done for each thread, since the same workload is split across more cores. This is true especially when the CPU-cores on top of which Metronome threads run are shared with other workload. In fact, the multi-thread approach reduces the per-CPU load of Metronome. 
This phenomenon of {\em resiliency} to the interference by other workloads will be assessed quantitatively in Section \ref{sec:impact}, along with the benefits for the applications sharing the same cores with Metronome.

Overall, with Metronome we propose an architecture where Rx queues can be efficiently shared among multiple threads: to each queue corresponds a lock which grants access 
to that queue. Threads can acquire access to a queue through our custom \texttt{trylock()} implementation, which provides non-blocking and minimal latency synchronization among them. For each of its queues, every thread tries to acquire the corresponding lock, and passes to the next queue if lock acquisition fails. Otherwise, if the thread wins the lock race it processes that queue as long as there are still incoming packets, then it releases the lock once the queue is idle. Once a thread has processed (or at least has tried to process) the Rx queues, it can go to sleep for a period of time proportional to (and controllable in a precise fine-grain manner depending on) the traffic weight it has experienced during its processing. Scheduling an awake-timeout through a fine-grained sleep service enables very precise and cheap thread-sleep periods, which are essential at 10Gb+ rates, and can still provide resource savings at lower rates. How a thread can elicit an awake-timeout period without incurring an Rx queue filling is carefully explained through our analytical model in Section \ref{sec:model}. This model is used to make the Metronome architecture self-tune its operations, providing suited trade-offs between resource usage (CPU cycles and energy) and packet processing performance.


\section{Metronome Adaptive tuning} 
\label{sec:model}
In this section we provide an approach to adaptively tune the behavior of the Metronome architecture. Metronome is designed to operate via a sequence of {\em renewal cycles} $\Theta(i)$, which alternate {\em Vacation Periods} with {\em Busy Periods}. As shown in Figure \ref{fig:cycle}, a {\em vacation period} $V(i)$ is a time interval where all the deployed packet-retrieval threads are set to {\em sleep mode}, hence incoming packets, labeled as $N_V(i)$ in the figure, get temporarily accumulated in the receive  buffer. For simplicity, we first consider a single Rx queue, then we expand our model to multiple queues in Section \ref{sec:multi}. When the first among the sleeping threads wakes up and wins the race, via a successful {\tt trylock()}, for handling the incoming packets from the Rx queue, a {\em busy period} $B(i)$ starts. This period will last until the whole queue is depleted by either the $N_V(i)$ formerly accumulated packets, as well as the new $N_B(i)$ packets arriving along the busy period itself $B(i)$---see the example in Figure \ref{fig:cycle}.

After depleting the queue, the involved thread will return to sleep. Note that other concurrent threads which wake up during a busy period will have no effect on packet processing---failing in the {\tt trylock()} they will just note that Rx queue unloading is already in progress and will therefore instantly return to sleep, thus freeing CPU resources for other tasks. 

\subsection{Metronome Multi-Threading Strategy}
\label{s4:metro}

As later demonstrated in Section \ref{sec:impact}, Metronome relies on multiple threads to guarantee increased robustness against CPU-reschedule delays of each individual Metronome thread, which is no longer in sleep state---the sleep timeout has fired and the thread was brought onto the OS run-queue. Such delay can be caused by CPU-scheduling decisions made by the OS---we recall that these decisions depend on the thread workload, their relative priorities and their current binding towards CPU-cores.

In such conditions, Metronome's control of the vacation period duration is not direct, as it would be in the single-thread case by setting the relevant timer, but it is {\em indirect} and stochastic, as this period is the time elapsing between the end of a previous busy period and the time in which some deployed thread awakes again and acquires the role of manager of the Rx queue. The question therefore is: {\em how to configure the awake timeouts of the different deployed Metronome threads}? 

Unfortunately, the simplest possible approach of {\em equal} timeouts comes along with performance drawbacks: we will demonstrate later on (see Figure \ref{fig:testtl}) that when timeouts are all set to a same value, CPU consumption significantly degrades as load increases, which is antithetic with respect to the objectives of Metronome. Indeed, especially under heavy packet arrival rate, threads would wake-up, therefore consuming CPU cycles, just to find out that another thread is already doing the job of unloading the Rx queue.

\begin{figure}[t]
    \centering
	\includegraphics[width=0.45\textwidth]{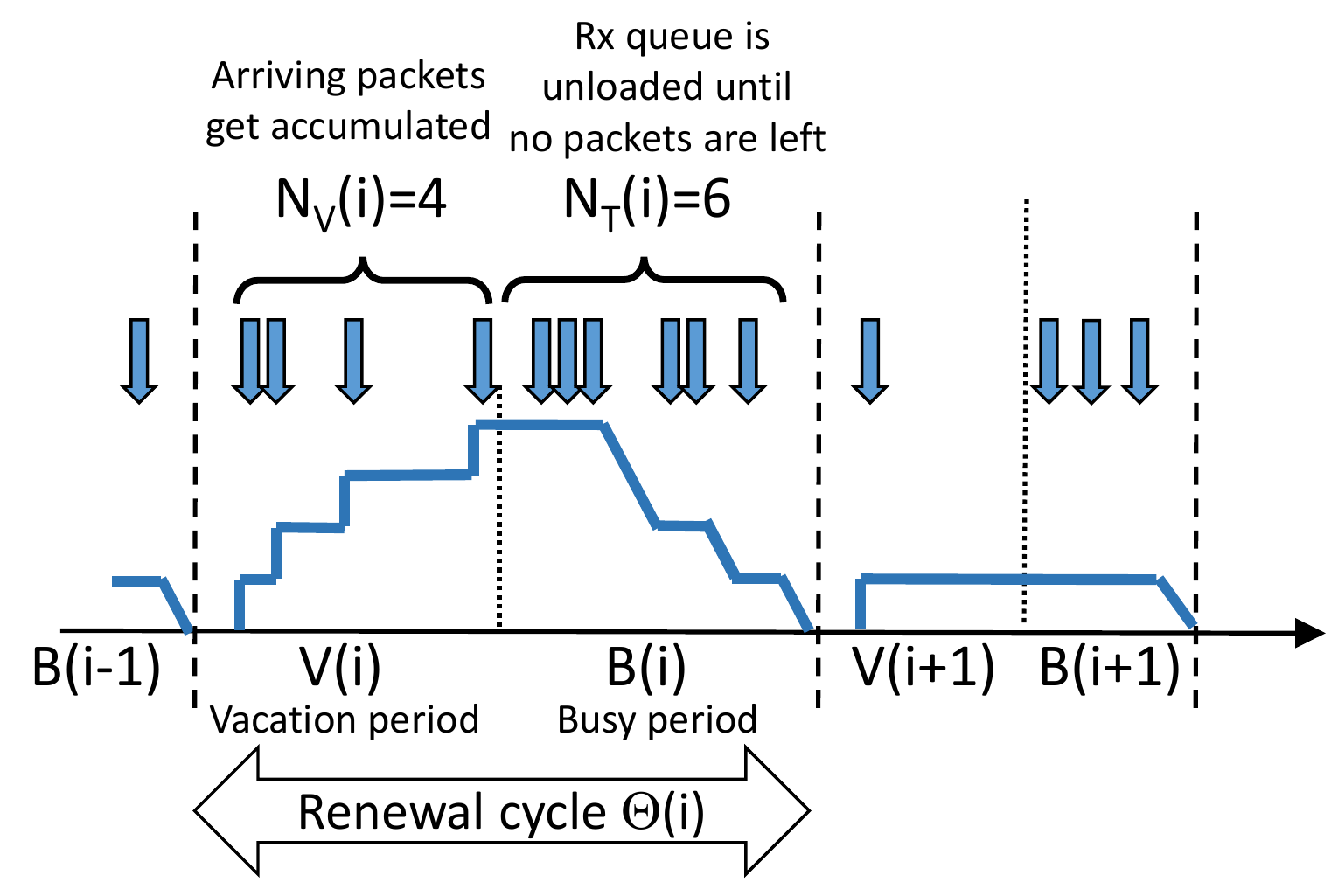}
	\vspace{-0.4cm}
	\caption{System model \& renewal cycle}
	\label{fig:cycle}
    \centering
	\vspace{0.2cm}
	\includegraphics[width=0.45\textwidth]{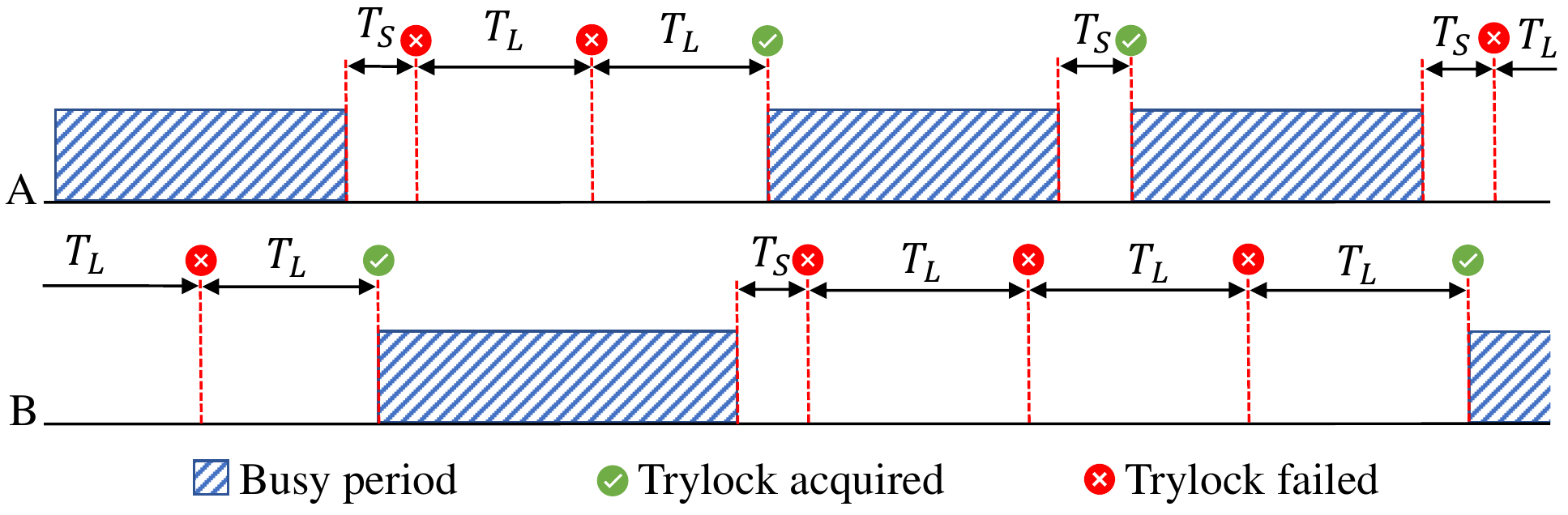}
	\caption{Vacation period and timeline of residual awake timeouts}
	\label{fig:wakeup}
	\vspace{-10pt}
\end{figure}

We thus propose a {\em \bf diversity-based} strategy for configuring the wake-up timeouts of different threads, which aims at mimicking a classical {\em primary/backup} approach, but without any explicit 
(and necessarily adding some extra CPU consumption) 
coordination, i.e. by using purely random access means. Each thread independently classifies itself as being in {\em primary} or {\em backup} state, according to the following rules:  
\begin{itemize}
    \item A thread becomes {\em primary} when it gets involved in a service time (it is the winner of the {\tt trylock()} based race); at the end of the busy period it carried out, it reschedules its next wake-up time after a "short" time interval $T_S$;
    \item A thread classifies itself as {\em backup} when it wakes-up and finds an on going busy period (i.e. another thread is already unloading the queue); it then schedules its next wake-up time after a "long" time interval $T_L>T_S$.  
\end{itemize}
In high load conditions, the above rules yield a scenario in which one thread at a time (randomly changing in the long term----see Figure \ref{fig:wakeup}) is in charge to poll the Rx queue at a reasonable frequency, whereas all the remaining ones occasionally wake up just for fall-back acquisition of the ownership on the Rx queue if for some reason the thread that was primary gets delayed, e.g. by the OS CPU-scheduling choices. Conversely, at low loads more threads will happen to be simultaneously in the {\em primary} state, thus permitting to significantly relax the requirements on the "short" awake timeout $T_S$ and motivating the adaptive strategy introduced in  Section \ref{s4:dynamic}.
\subsection{Metronome Analysis}
\label{s4:perf}

\subsubsection{Background}
Let us non-restrictively assume that, once a thread wakes up, the packets accumulated in the Rx buffer get retrieved at a constant rate $\mu$ packets/seconds (this assumption is discussed in more details in Appendix II). It readily follows that the duration of the busy period $B(i)$ depends on the number of accumulated packets, and, more precisely, it comprises two components: i) the time needed to deplete the first $N_V(i)$ packets arrived during $V(i)$, plus ii) the extra time needed to deplete the next $N_B(i)$ packets arrived since the start of the the busy period---in formulae:
\begin{equation}
B(i) = \frac{N_V(i)+ N_B(i)}{\mu} 
\label{eq:ti}
\end{equation}
Since $N_V(i)$ and $N_B(i)$ depend on the vacation period $V(i)$, in most generality drawn from a random variable $V$, we can take conditional expectation at both sides of  (\ref{eq:ti}) with respect to $V$.  Being $\lambda$ the (unknown) mean packet arrival rate, we obtain the following fixed point equation\footnote{In the derivation, we exploited the following well known fact (direct consequence of the Little's Result): the average number $E[N]$ of packets arriving during a time interval of mean length  $E[T]$ is $E[N]=\lambda E[T]$.} in $E\left[B | V\right]$: 
\begin{equation}
E\left[B | V\right] = \frac{1}{\mu} E\left[N_V(i)+ N_B(i) | V \right] = \frac{\lambda}{\mu} \left(V + E\left[B | V \right]\right).
\label{eq:ti:vi}
\end{equation}
which yields an explicit expression of how a busy period  $E\left[B | V\right]$ is affected by the relevant vacation period:
\begin{equation}
E\left[B | V\right] = V \frac{\lambda/\mu}{1-\lambda/\mu}
\label{eq:bv}
\end{equation}
If we conveniently define $\rho=\lambda/\mu$, we can derive an explicit expression which relates $\rho$ to the controllable Vacation Period duration $V$ and the relevant observable Busy Period $E[B|V]$--- this expression will be indeed used to estimate $\rho$ in Section \ref{s4:dynamic}:
\begin{equation}
\rho = \frac{E\left[B | V\right]}{V+E\left[B | V\right]}
\label{eq:rho}
\end{equation}

\begin{figure*}[tb]
\centering
\begin{minipage}{0.31\textwidth}
  \centering
\includegraphics[width=\textwidth]{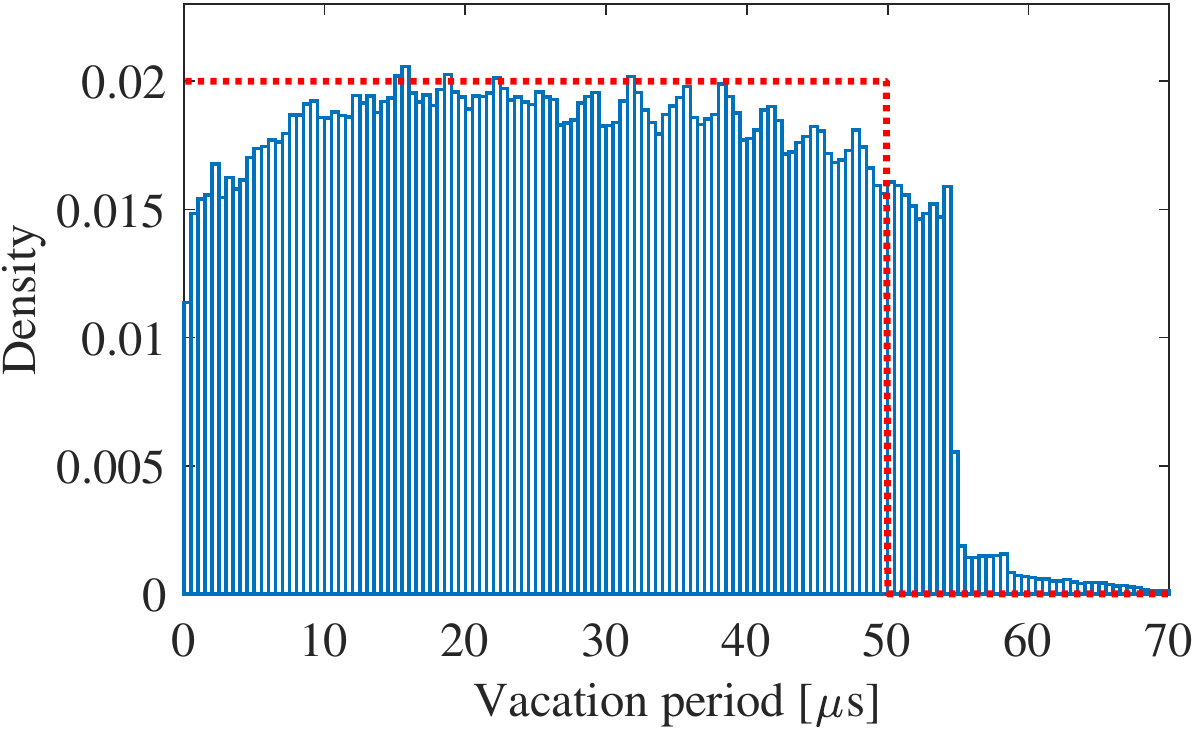}
\subcaption{2 cores}
\end{minipage}%
\begin{minipage}{0.31\textwidth}
  \centering
\includegraphics[width=\textwidth]{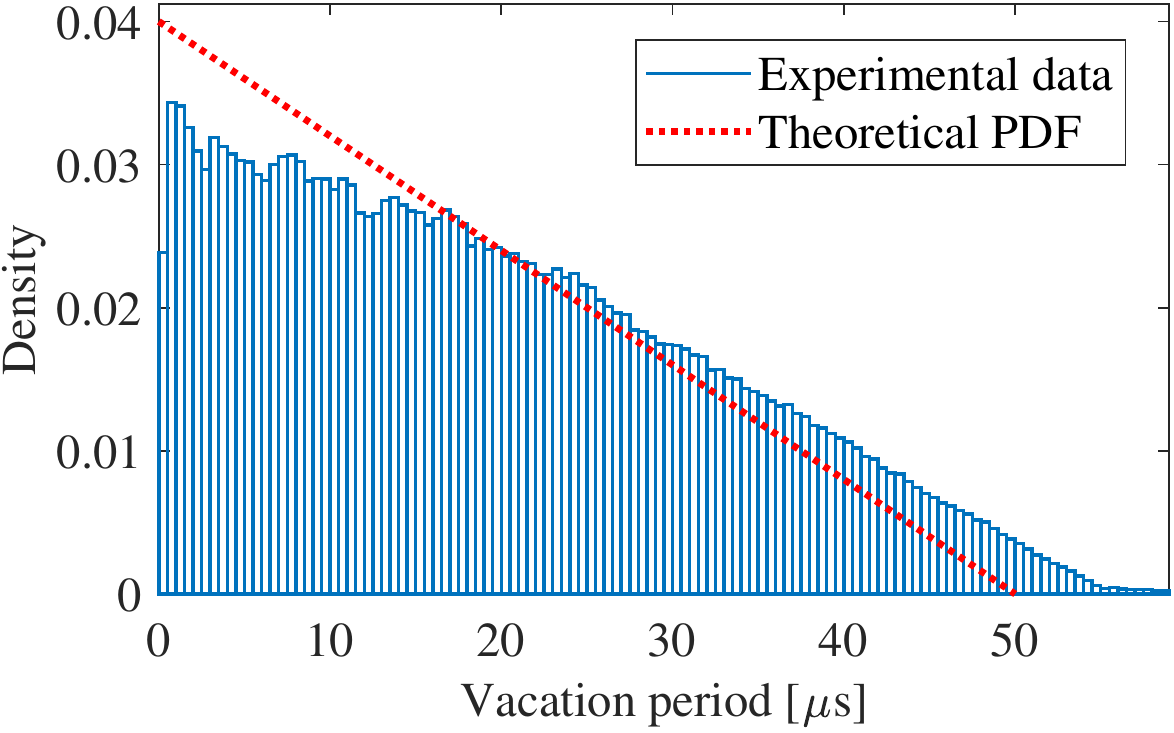}
\subcaption{3 cores}
\end{minipage}%
\begin{minipage}{0.31\textwidth}
  \centering
\includegraphics[width=\textwidth]{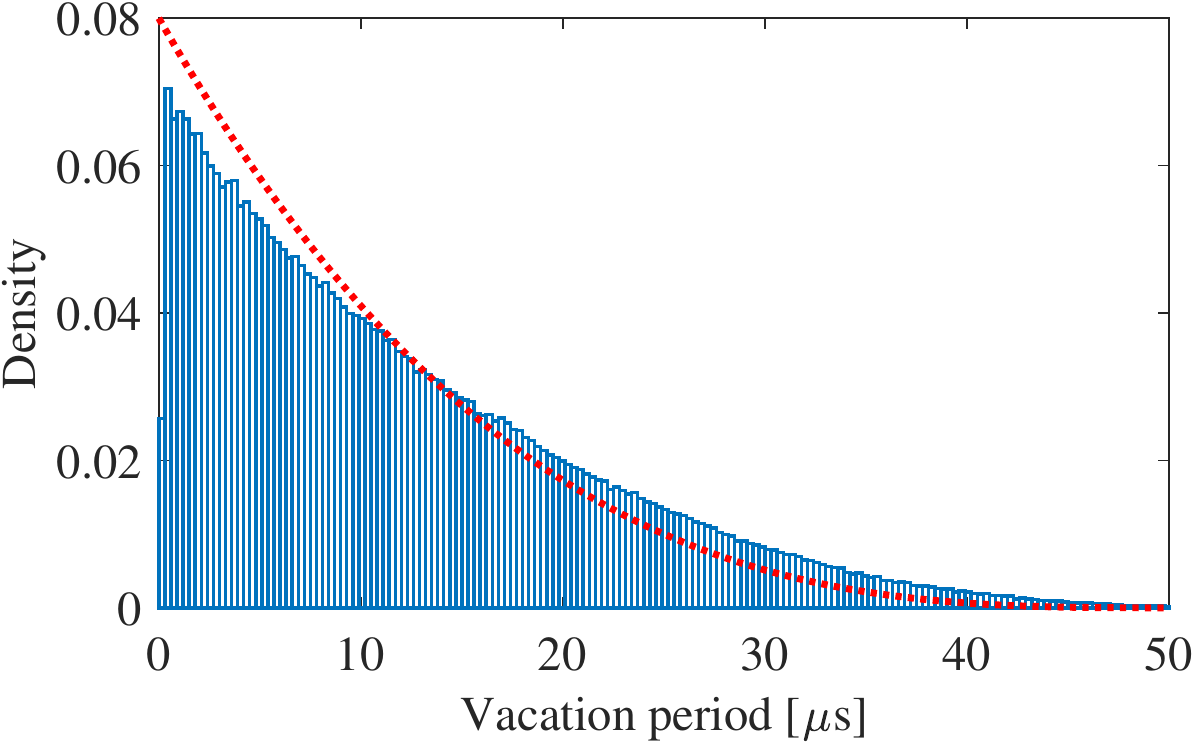}
\subcaption{5 cores}
\end{minipage}
\caption{Vacation period PDF: analysis vs experiments, $T_S=T_L$} \label{fig:pdf}
\end{figure*}

\subsubsection{Vacation Period statistics at high load}
It is useful to start from two simplified mean-value analyses relying on two opposite sets of assumptions valid at either high load or low load. The two different models will be then blended into a single one in Section \ref{s4:dynamic}. Let $M \geq 2$ be the number of deployed Metronome threads. In {\em high load conditions}, for reasons that will soon become evident, we can assume that only one of such threads is in the primary state, whereas all the remaining $M-1$ are in backup state. Once the primary thread releases the Rx queue lock and schedules its short timer $T_S$, two possible cases may occur:
\begin{itemize}
    \item no backup thread wakes up during the sleep timeout $T_S$; in this case the primary thread will get back control of the Rx queue for the next round, and will remain primary;
    \item one of the remaining $M-1$ backup threads wake up {\em before} the end of the sleep timeout $T_S$ and thus becomes primary; when the former primary thread wakes up, it will find a busy period\footnote{In high load conditions, owing to equation \ref{eq:bv}, the average busy period lasts significantly longer than the vacation period.} and will therefore acquire the role of backup thread, rescheduling its next wake up timeout after a time $T_L$.
\end{itemize}


Let us now make the assumption that the (current) $M-1$ backup threads were earlier CPU-rescheduled at independent random times. This {\em Decorrelation} assumption, indeed later on verified in Figure \ref{fig:pdf} using experimental results, is justified by the fact that each service time, due to its random duration, de-synchronizes the primary thread CPU-reschedule from the remaining ones; since after a few busy cycles all threads will have the chance to become primary, even if initially being CPU-scheduled at around the same times, their CPU-rescheduling instants will rapidly "decorrelate".

The statistics of the random variable $V$ (vacation period) can be computed as the {\em minimum} between i) the fixed wake-up timeout $T_S$ of the primary thread, and ii) the wake-up timeout of any of the remaining $M-1$ threads, which, owing to the previous decorrelation assumption, have been CPU-rescheduled in any random instant in the range between $0$ and $T_L$ before the end of the current busy period. It readily follows that the cumulative probability distribution function of $V$ is given by:
\begin{equation}
    CDF_V(x) = P(V\leq x) = \begin{cases} 1-\Big(1-\frac{x}{T_L}\Big)^{M-1} & x<T_S \\
    1 & x\geq T_S \end{cases}
    \label{eq:cdf}
\end{equation}
and the mean vacation period for a given configuration of the short and long awake timeouts, and for a given number of threads, is trivially computed as: 
\begin{equation}
E[V] = \int_0^{T_S} (1-CDF_V(x)) d \! x =  \frac{T_L}{M} 
\left(1 - \left(1-\frac{T_S}{T_L}\right)^M\right)
    \label{eq:ev}
\end{equation}
Finally, the probability that one of the $M-1$ backup threads gains access to the Rx queue at its wake-up time is given by:
\begin{equation}
P_{s,succ} = \int_0^{T_S} \frac{1}{T_L} \left(1-\frac{x}{T_L}\right)^{M-2} d \! x = \frac{\left(1-\frac{T_S}{T_L}\right)^{M-1}}{M-1} 
\end{equation}


\subsubsection{Vacation period statistics at low load}
While, at high load, a neat pattern emerges in terms of one single primary thread at any time, with multiple backup threads, it is interesting to note that at low load Metronome yields a completely different behavior. Indeed, owing to equation (\ref{eq:bv}), as the offered load reduces, the average busy period duration becomes small with respect to the vacation period duration. It follows that when a primary thread gets control of the Rx queue, it very rapidly releases such control, so that another thread waking up will find the queue available with high probability. It follows that in the extreme case, {\em all} threads will always remain in the primary state\footnote{This is because each time an awaken thread finds the Rx queue not locked by another thread, then it acquires the primary role thanks to its successful {\tt trylock()} operation.} and thus will periodically reschedule their next wake-up times after a  short interval $T_S$. This case is even simpler to analyze than the previous one, as the CDF of the vacation time directly follows from (\ref{eq:cdf}) by simply setting $T_L=T_S$ and by considering $M$ ``competitors'', in formulae:
\begin{equation}
    CDF_V(x) = P(V\leq x) = 1-\Big(1-\frac{x}{T_S}\Big)^M 
    \label{eq:cdf:low}
\end{equation}
and mean vacation period simplifying to $E[V] = T_S/M$.

\subsubsection{Experimental verification of the decorrelation assumption}
To verify the validity of the decorrelation assumption used in the above models, Figure \ref{fig:pdf} compares the probability distribution function obtained by taking derivative of the CDF in equation (\ref{eq:cdf}), i.e., for $x<T_S$,
\begin{equation}
    PDF_V(x) = \frac{M-1}{T_L} \Big(1-\frac{x}{T_L}\Big)^{M-2}
    \label{eq:pdf}
\end{equation}
with experimental results. We have specifically focused on the case $T_L=T_S$ as in this case the formula in equation (\ref{eq:cdf}) is expected to hold independently of the load (primary and backup threads use the same awake timeouts). Results, obtained with awake timeouts set to $50 \mu s$ and different numbers of threads $M$, suggest that the decorrelation approximation is more than reasonable and the proposed model is quite accurate. Furthermore, the results also show that, in the real case---although rarely---actual CPU-reschedules after a sleep period can occur after the maximum time delay $T_L$, because of CPU-scheduling decisions by the OS---for example favoring OS-kernel demons.
However, such impact becomes almost negligible in Metronome with just $M=3$ deployed threads, pointing to the relevance of the adopted multi-threading approach.

\subsection{Adaptation policy under general load conditions}
\label{app:math}
We propose a simplified, but still theoretically motivated, approach which allows us to blend the results obtained via the two extreme low and high load models into a single and convenient analytical framework. 

More specifically, in {\em intermediate load conditions} we cannot anymore assume that just one single thread (as in high load conditions), or all threads (as in low load conditions), are in primary state along time. Rather, a part from the {\em single} thread that has last depleted the Rx queue, which is therefore surely in primary state, also {\em some} of the remaining $M-1$ threads will be in primary state whereas others will be in backup state. Let us therefore introduce a random variable $P$ which represents the number of the remaining threads in primary state. $M-1-P$ will therefore be the number of remaining threads in secondary state. 

Let us now assume that each of the remaining $M-1$ threads can be {\em independently} found in primary or backup state with probability $p$ (which will be determined later on). Then, the random variable $P$ representing the number of remaining threads in primary state trivially follows the Binomial distribution:
\[\mathrm{Prob}(P=k) = \binom{M \! - \! 1}{k} p^k (1-p)^{M-1-k} \]
Then, we can compute the average vacation time also in intermediate load conditions, by taking conditional expectation over this newly defined random variable $P$. This permits us to generalize equation (\ref{eq:ev}) as follows:
\[ E[V] = E[E[V|P]] = \]
\[ = \sum_{k=0}^{M-1} \binom{M \! - \! 1}{k} p^k (1-p)^{M-1-k}
\int_0^{T_S} \!\! \Big(1-\frac{x}{T_S}\Big)^{k} \cdot\]
\[\cdot \Big(1-\frac{x}{T_L}\Big)^{M-1-k} d \! x = \]
\[ = \int_0^{T_S} \left(1-\frac{p x}{T_S} - \frac{(1-p) x}{T_L}\right)^{M-1} d\!x = \]
\[ =\frac{1-\left((1-p)(1-T_S/T_L)\right)^M}{M\left(\frac{p}{T_L} + \frac{1-p}{T_S}\right)} \]
Furthermore, assuming  $T_L>>T_S$, we can conveniently simplify the above expression and approximate it as:
\begin{equation}
   E[V] =\frac{T_S}{M}\cdot \frac{1-(1-p)^M}{p}
   \label{eq:ev:p}
\end{equation}
Note that, for $p\rightarrow 0$, namely when the probability to find another thread in the primary state becomes zero (high load conditions), equation (\ref{eq:ev:p}) converges to the expected value $T_S$, whereas $E[V] = T_S/M$ for $p=1$ (as for low load conditions, i.e. all the threads becoming primary). 

As a last step, it suffices to relate $p$ with the offered load. To this purpose, let $\rho = \lambda/\mu$ be the probability that the Rx queue is busy at a random sample instant. It is intuitive to set $p=(1-\rho)$, as the probability $p$ that a thread is in the primary state is the probability that when this thread has last sampled the queue, it has found it idle, i.e. $1-\rho$. This finally permits us to formally support our proposed formula (\ref{eq:ts}) as the load-adaptive $T_S$ setting strategy. Summarizing for the reader's convenience, being $\bar V$ a constant target vacation period, and $\rho$ the current load estimate, $T_S$ can be set as:
\[ T_S = M \frac{1-\rho}{1-\rho^M}  \cdot {\bar V} \]
Note that this rule can be conveniently rewritten in a more intuitive and simpler to compute form, as:
\[T_S = M \frac{1-\rho}{1-\rho^M} = {\bar V} \frac{M}{1+\rho + \cdots + \rho^{M-1}}\]

\subsection{Metronome Adaptation and  Tradeoffs}
\label{s4:dynamic}
Whenever the mean arrival rate is non-stationary, but  varies at a time scale reasonably longer than the cycle time, the load conditions can be trivially run-time estimated using equation (\ref{eq:rho}). For instance, the simplest possible approach is to consider for $\rho(i)=\lambda(i)/\mu$ the exponentially weighted estimator:
\begin{equation}
\rho(i) = (1-\alpha) \rho(i-1) + 
\alpha \frac{B(i)}{V(i)+B(i)}
\label{eq:est}
\end{equation}

Established that measuring the load is not a concern for Metronome, a more interesting question is to devise a mechanism which adapts the awake timeouts to the time-varying load. 
The obvious emerging trade-off consists in trading the polling frequency, namely the frequency at which threads wake up, with the duration of the vacation period which directly affects the packet latency. Indeed, if we assume that the serving thread is capable to drain packets from the Rx queue at a rate $\mu$ greater than or equal to the link rate, namely the maximum rate at which packets may arrive (in our single-queue experiments, 10 gigabit/s), then once the thread starts the service, packets will no longer accumulate delay. Therefore, the worst case latency occurs when a packet arrives right after the end of the last service period, and is delayed for an entire vacation period. 

It follows that an adaptation strategy that {\em targets a constant vacation period duration irrespective of the load} appears to be a quite natural approach. Let us recall that, under the assumption $T_L>>T_S$, the average vacation period at high load given by equation (\ref{eq:ev}) simplifies to $E[V] \approx T_S$. Conversely, at low load, we obtained $E[V]=T_S/M$. Therefore, being $\bar V$ our target constant vacation period, the rule to set the timer $T_S$ at either high or low loads reduces to:
\begin{equation}
\begin{cases} 
T_S = {\bar V} & \mathrm{high load} \\
T_S = M \cdot {\bar V} & \mathrm{low load}
\end{cases}
\label{eq:adapt:hl}
\end{equation}

The analysis of the general case (intermediate load) is less straightforward, but can be still formally dealt with by assuming that threads are independent and are in primary or backup state according to the probability that, while they wake up, they find the Rx queue idle or busy, respectively. As shown in Section \ref{app:math}, we can prove that, in this general case, under the assumption $T_L>>T_S$, the rule to set the timer $T_S$ becomes:
\begin{equation}
\label{eq:ts}
T_S = M \frac{1-\rho}{1-\rho^M}  \cdot {\bar V} 
\end{equation}
which, as expected, converges to (\ref{eq:adapt:hl}) for the extreme high load case $\rho \rightarrow 1$ and the extreme low load case $\rho \rightarrow 0$.

Finally, we stress that Metronome does {\em not} sacrifice latency, but provides the {\em possibility to trade latency for CPU consumption}. Indeed, the duration of the chosen vacation period will determine the performance/efficiency trade-off: the longer the chosen vacation time, the lower the polling rate and thus the CPU consumption, at the price of a higher latency. If a deployment must guarantee low latency then it should either configure a small vacation time target, or even disable Metronome and use standard DPDK. 

\subsection{The multiqueue case}\label{sec:multi}
When Metronome is used with 40+Gb NICs, one queue becomes not enough to sustain line rate traffic. Therefore, a split of the incoming traffic into multiple receive queues through RSS is needed. We now introduce the $N$ parameter, which represents the number of Rx queues for a certain NIC. Given $M$ as the total number of threads in the system, we believe it should be at least as big as $N$, so that every queue can have one primary thread associated to it ($M\geq N$). In this scenario, we have $N$ primary threads (since everyone of them has won the lock race for a different queue) and $M-N$ secondary threads. In a scenario with multiple queues, we believe it is not efficient to statically bind a thread to a certain queue, so we propose a different approach:
\begin{itemize}
    \item once a \textbf{primary} thread has won the race and depleted a queue, it goes to sleep for a $T_S$ period and when it wakes up, it contends for the same queue as we know it is likely for it to win the race again.
    \item once a \textbf{backup} thread has lost a lock race, it chooses the queue to be contended at its next wakeup randomly.
\end{itemize}
The random selection of the next queue for the backup thread ensures us a certain decorrelation among the threads in the next queue selection and also fairness with respect to the queue checks. While the $T_L$ value remains fixed, we update equation (\ref{eq:ts}) as follows:
\begin{equation}
\label{eq:ts_multi}
T_S = \frac{M}{N} \cdot \frac{1-\rho_i}{1-\rho_i^\frac{M}{N}}  \cdot {\bar V}\;\;\;\text{for} \;i=1,\dots,N
\end{equation}
We notice two differences with the single queue version. The former is that the $M$ parameter is now replaced with $M/N$, as that is the average number of threads taking care of a certain queue at any moment. The latter is that the $\rho$ parameter is now per-queue based, as each queue can experiment different traffic rates (and therefore, queue occupancy) at any time.
\section{Experimental results} \label{sec:eval}
Our experimental campaign starts with the appropriate tuning for the ${\bar V}$, {$T_L$} and M parameters and the analysis of the subsequent tradeoffs. We then test the adaptation capabilities of Metronome in Section \ref{sec:adaptation}. Section \ref{sec_comparison} discusses in detail both strengths and weaknesses of Metronome and static DPDK in different aspects (latency, CPU usage and power consumption). Section \ref{sec:xdp} compares Metronome and XDP, while Section \ref{sec:impact} shows the impact of Metronome in common CPU sharing scenarios. While tests up to Section \ref{sec:impact} have been conducted with a single Rx queue 
(using Intel X520 NICs), Section \ref{sec:eval_multi} evaluates Metronome in a multi queue scenario (with Intel XL710s).
For evaluating the system we used a server running Linux kernel 5.4 equipped with Intel\textregistered\ Xeon\textregistered\ Silver @2.1 GHz, running the \texttt{l3fwd} DPDK application \cite{l3fwd} on an isolated NUMA node and generating traffic with MoonGen \cite{moongen}. For benchmarking our system, we used the evaluation suite provided by Zhang et al. in \cite{softwareswitches}, as well as the Intel RAPL package \cite{rapl} and the \texttt{getrusage()} syscall to retrieve energy usage and CPU consumption. 
Tests are done with 64B packets, as this is the worst case scenario\footnote{For tests regarding latency, since \cite{softwareswitches} uses Moongen's timestamping capabilities, it is necessary to add a 20B timestamp to the timestamped subset of packets, thus giving rise to a minimal difference in terms of offered throughput.}. Unless explicitly stated, the tests are executed using the \texttt{performance} CPU power governor and with parameters ${\bar V}$= 10 $\mu$s, ${T_L}$= 500 $\mu$s, M=3---each choice is motivated in the following section. Further tests for two different applications are also shown (see Figure \ref{fig:cpu:ipsec_flowatch}).

\subsection{Parameters Tuning} \label{sec:tuning}
\begin{figure}[tb]
    \centering
    \subfloat[][10Gbps traffic]{%
        \includegraphics[width=0.52\linewidth]{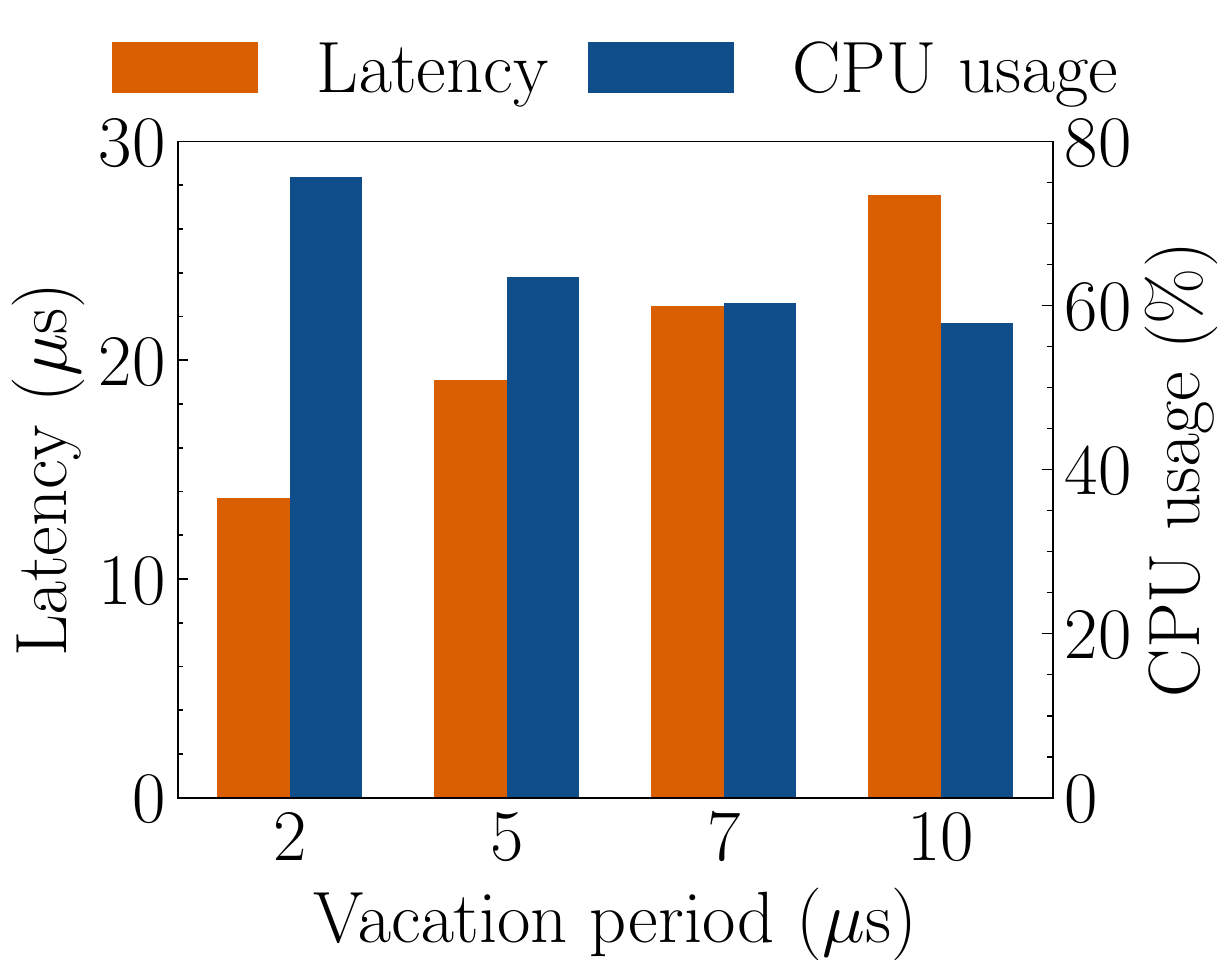}%
        \label{fig:a}%
        }%
    \subfloat[][5Gbps traffic]{%
        \includegraphics[width=0.52\linewidth]{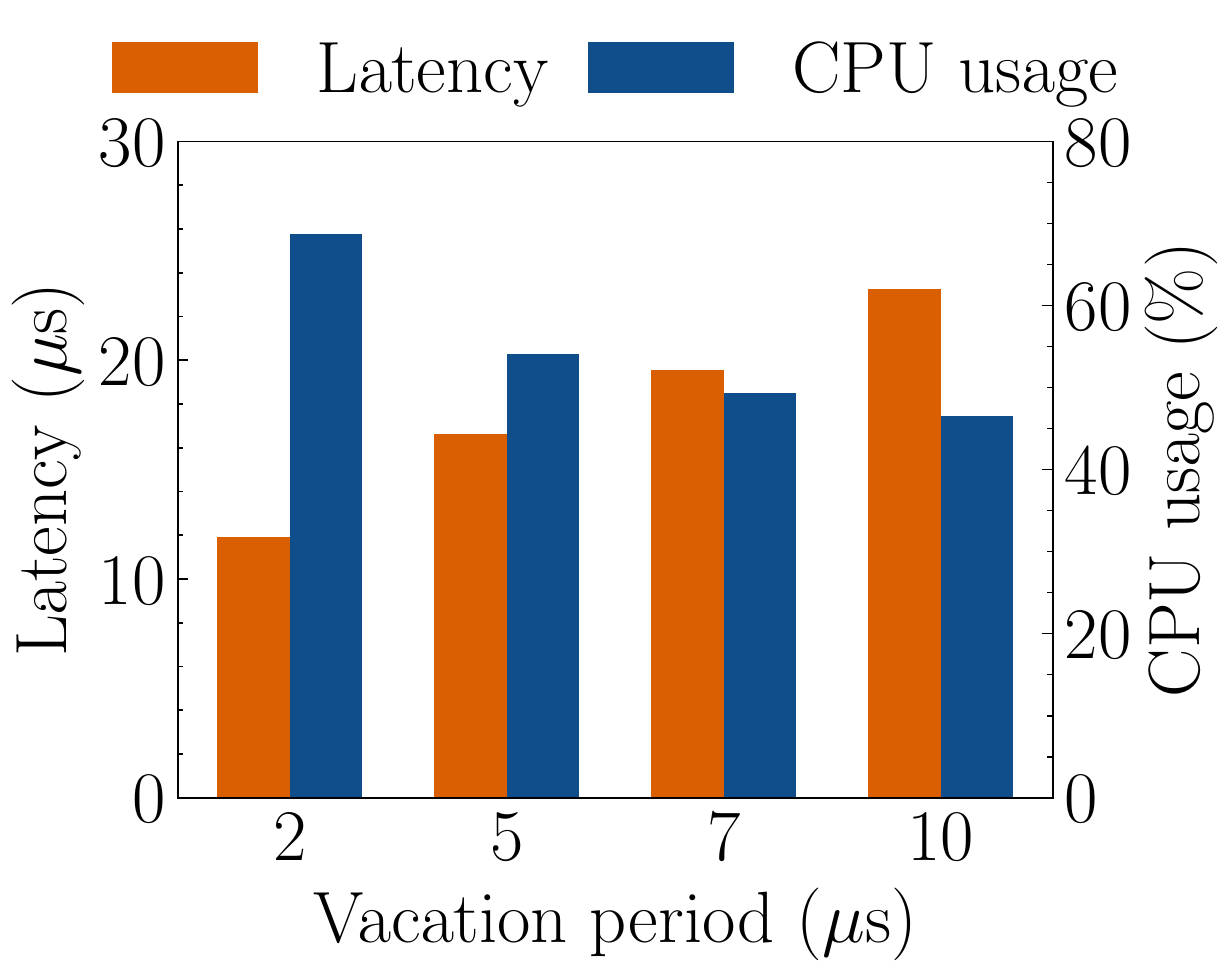}%
        \label{fig:b}%
        }%
    \caption{Latency and CPU usage for different different target vacant times.} \label{fig:vtuning}
\end{figure}

\begin{figure}[tb]
\vspace{-0.2cm}
\centering
\begin{minipage}[t]{.23\textwidth}
  \centering
  \includegraphics[width=.95\linewidth]{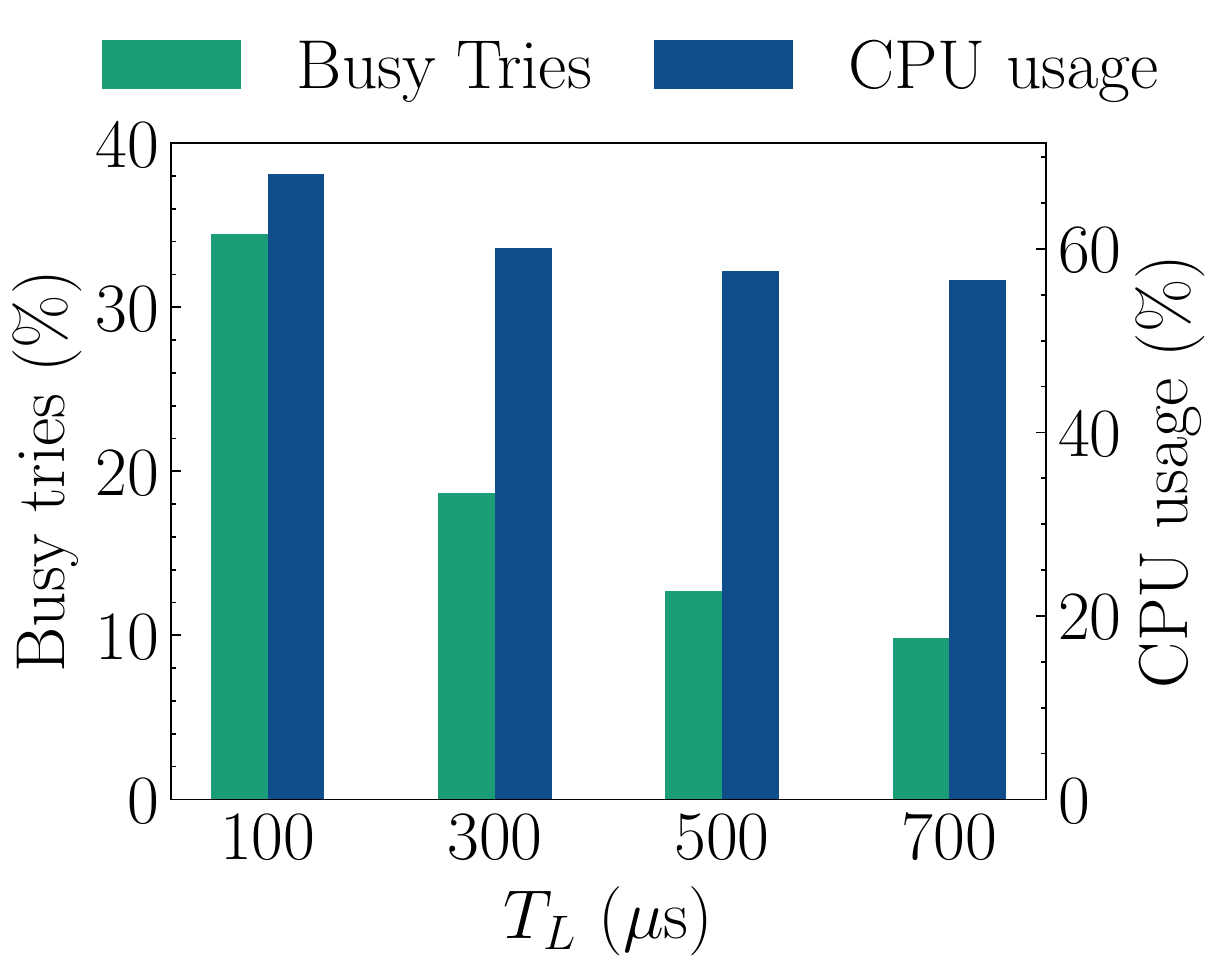}
  \captionof{figure}{Busy tries and CPU usage versus $T_L$.}
  \label{fig:testtl}
\end{minipage}\hfill
\begin{minipage}[t]{.23\textwidth}
  \centering
  \includegraphics[width=.95\linewidth]{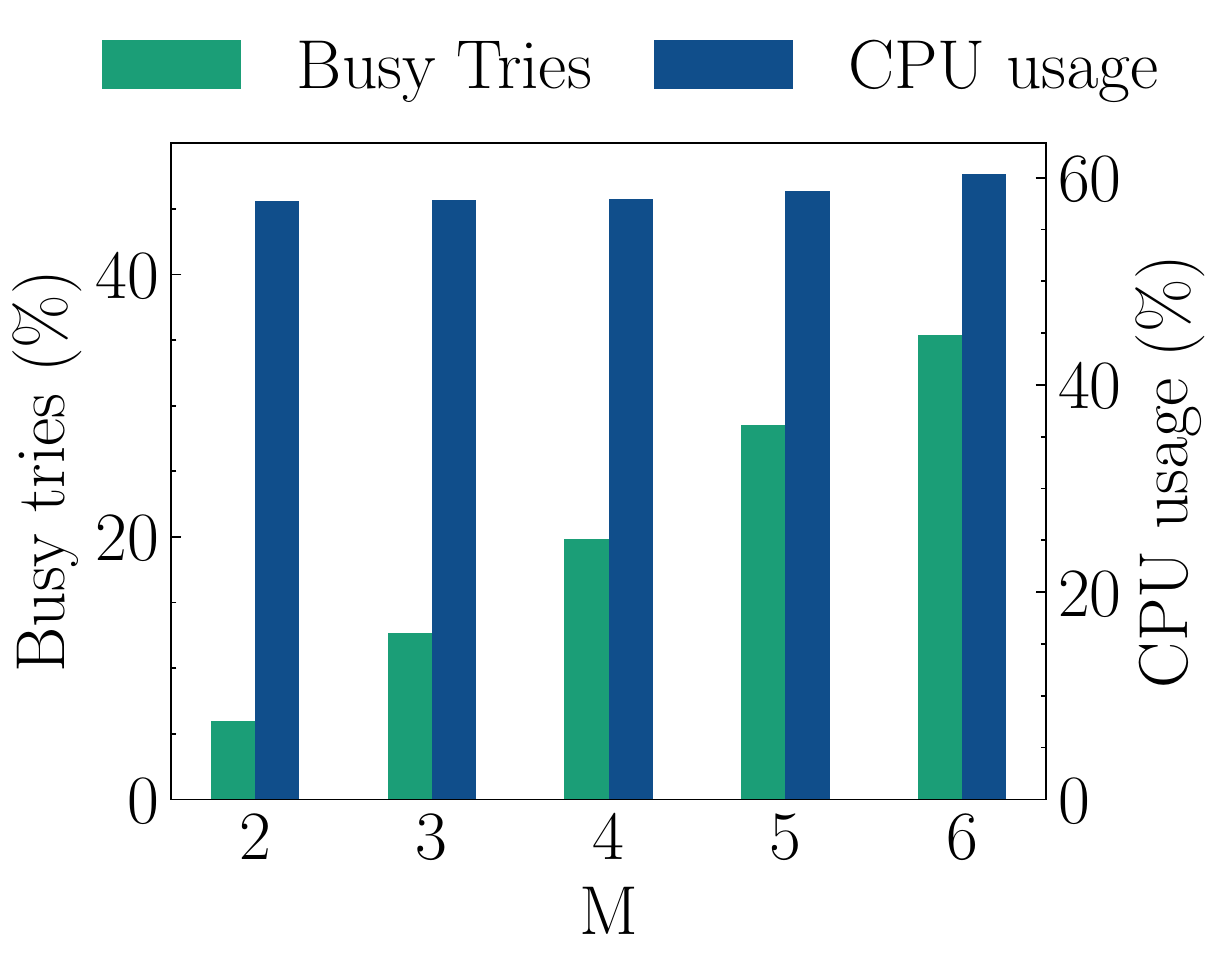}
  \captionof{figure}{Busy tries and CPU usage versus $M$.}
  \label{fig:threads}
\end{minipage}
\vspace{-10pt}
\end{figure}

\begin{figure}[hb]
    \centering
    \subfloat[10Gbps traffic]{%
        \includegraphics[width=0.47\linewidth]{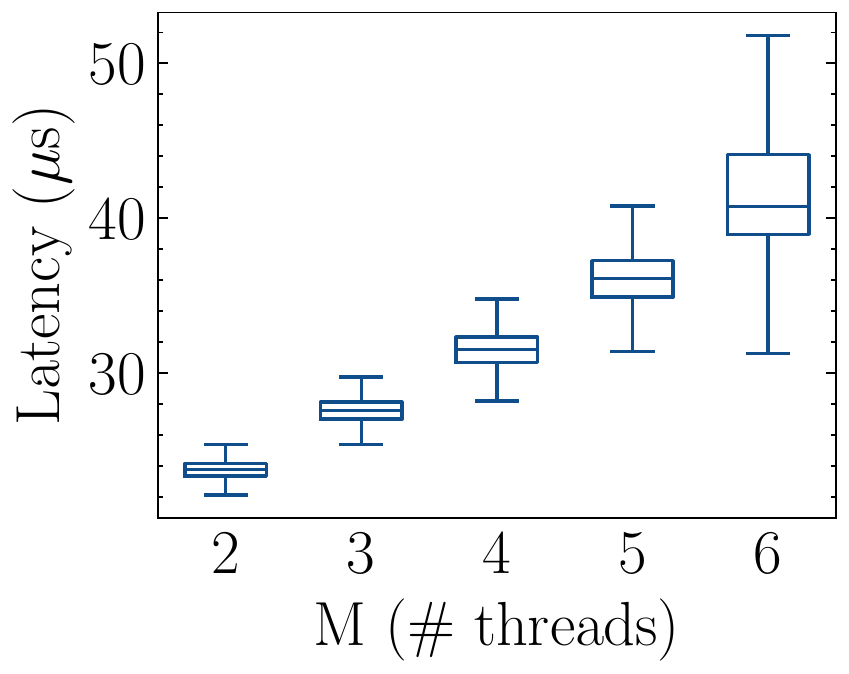}%
        \label{fig:threads10}%
        }%
        \hfill
        \subfloat[1Gbps traffic]{%
        \includegraphics[width=0.47\linewidth]{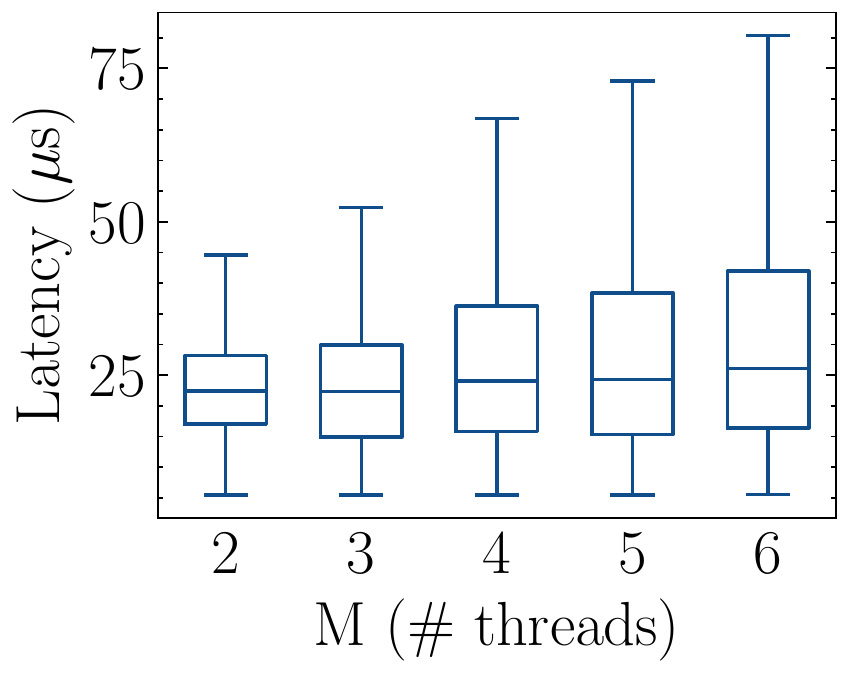}%
        \label{fig:threads1}%
        }%
    \caption{Latency vs. the number of threads M}\label{fig:threadslatency}
    \vspace{-0.4cm}
\end{figure}

\noindent 
First of all, we would like to find a vacation period ${\bar V}$ which permits us not to lose packets under line-rate conditions. Table \ref{tab:vac_time} shows packet loss, vacation period and busy period for different values of ${\bar V}$, which represents the target $V$ to be used when calling the {\tt hr\_sleep()} service: we found out that 10 $\mu$s is a good starting point as it provides no loss. The test was conducted using the suite's unidirectional p2p throughput test, as this test instantly increases the incoming rate from 0 to 14.88 Mpps, so as to be sure that this setting works even in the worst case scenario. We then analyzed the bidirectional throughput scenario by assigning 3 different threads to each Rx queue, as we found out that Metronome achieves the same maximum bidirectional throughput that DPDK can reach (11.61 Mpps per port) by constantly polling each Rx queue with a different thread.
\begin{table}[tb]
\scriptsize
\begin{tabular}{|c|c|c|c|c|}
\hline
Target V [$\mu s$] & Measured V [$\mu s$] & Measured B [$\mu s$] & $N_V$ & Loss ($\tcperthousand$) \\ \hline
5  & 11.67 & 13.40 & 172.39 & 0 \\ \hline
10 & 19.55 & 20.24 & 287.77 & 0 \\ \hline
12 & 21.99 & 22.86 & 326.30 & 0.0037 \\ \hline
15 & 26.23 & 27.25 & 385.18 & 0.023 \\ \hline
20 & 33.28 & 38.32 & 494.39 & 1.180 \\ \hline
\end{tabular}
\caption{Mean busy and vacation period, $N_V$ and packet loss for different target vacation periods.}
\vspace{-0.4cm}
\label{tab:vac_time}
\end{table}
Once a good suitable minimum value for ${\bar V}$ is found, we investigate how tuning ${\bar V}$ affects CPU usage and latency: indeed, as Table \ref{tab:vac_time} shows, the shorter ${\bar V}$, the less the queue is left unprocessed as the actual (namely, the measured) vacation time V decreases, so packets tend to experience a shorter queuing period. However, such an advantage does not come for free, as the CPU usage proportionally increases, as shown in Figure \ref{fig:vtuning} for different traffic volumes. We note that all these tests have been performed by relying on 3 Metronome threads.
\\
\noindent As for $T_L$, while letting backup threads sleep for a longer period of time alleviates the percentage of failed attempts  of  {\tt trylock()} (busy tries),  and therefore the number of wasted CPU cycles (as Figure \ref{fig:testtl} shows), a shorter $T_L$ means higher reactivity when the primary thread is interfered by OS CPU-scheduling choices. For our evaluation we chose 500 $\mu$s since (i) it is 50 times bigger than the maximum 
{$T_S$} possible value, we recall that our analytical model assumes that $T_L>>T_S$, (ii) Figure \ref{fig:testtl} shows that most of the advantage of increasing $T_L$ happens before 500 $\mu$s, while between 500 and 700 $\mu$s we experimented a difference of only 1\% in CPU usage and around 2\% in busy tries. 
As for M, the philosophy underlying Metronome is the one of exploiting multiple threads for managing a Rx queue, not the one using excessive (hence useless) thread-level parallelism. 
In fact, an excessive number of threads comes at almost no usefulness: Figure \ref{fig:threads} shows how the percentage of busy tries increases linearly with the number of threads, along with a slight cost increase in terms of CPU usage. Furthermore, increasing the threads number comes along with a significant cost in terms of latency, as the more the threads, the more frequently a primary thread switches to the backup role leading to longer sleep periods as stated in equation (\ref{eq:ts}). We experimented considerable latency implications especially at high rates, as Figure \ref{fig:threads10} shows. Even for much lower rates, a substantial increase in variance is still visible (see Figure \ref{fig:threads1}). 
By the above hints, the single-queue evaluation is done with 3 threads.


\subsection{Adaptation}\label{sec:adaptation}
\begin{figure}[t]
\vspace{-0.2cm}
    \centering
    \subfloat[Rate and $T_S$ estimation]{%
        \includegraphics[width=0.5\linewidth]{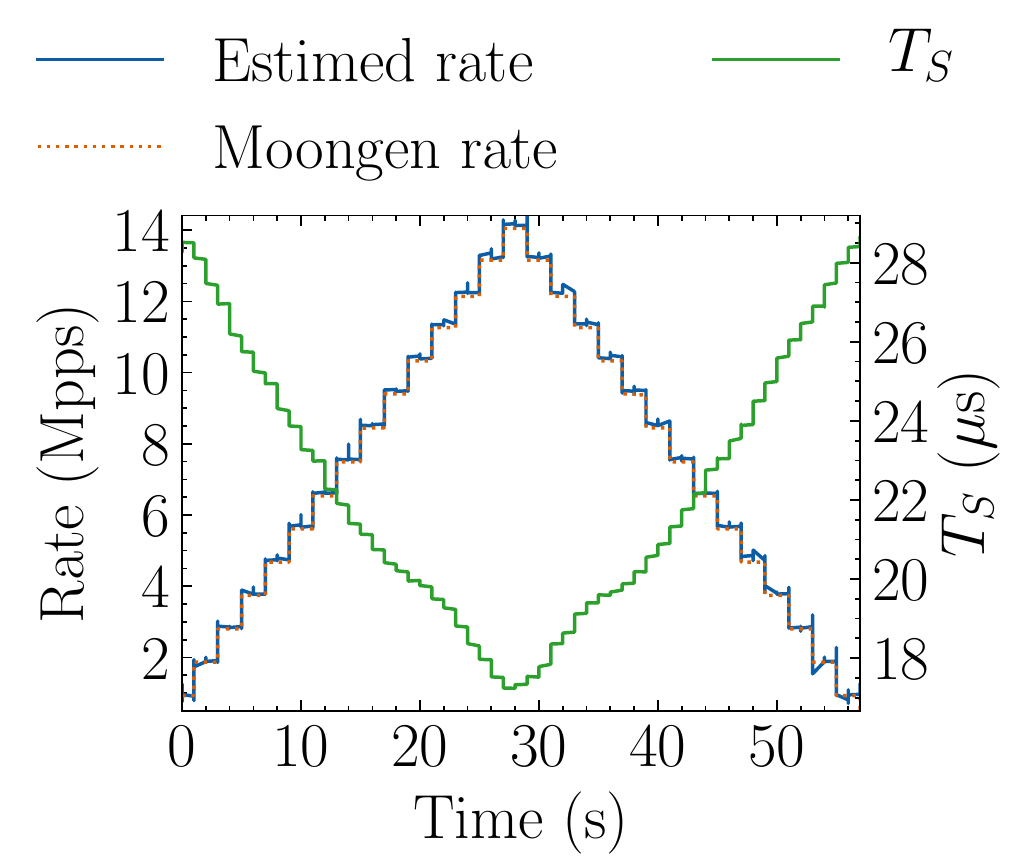}%
        \label{fig:adaptiverate}%
        }%
    \subfloat[CPU usage and $\rho$]{%
        \includegraphics[width=0.5\linewidth]{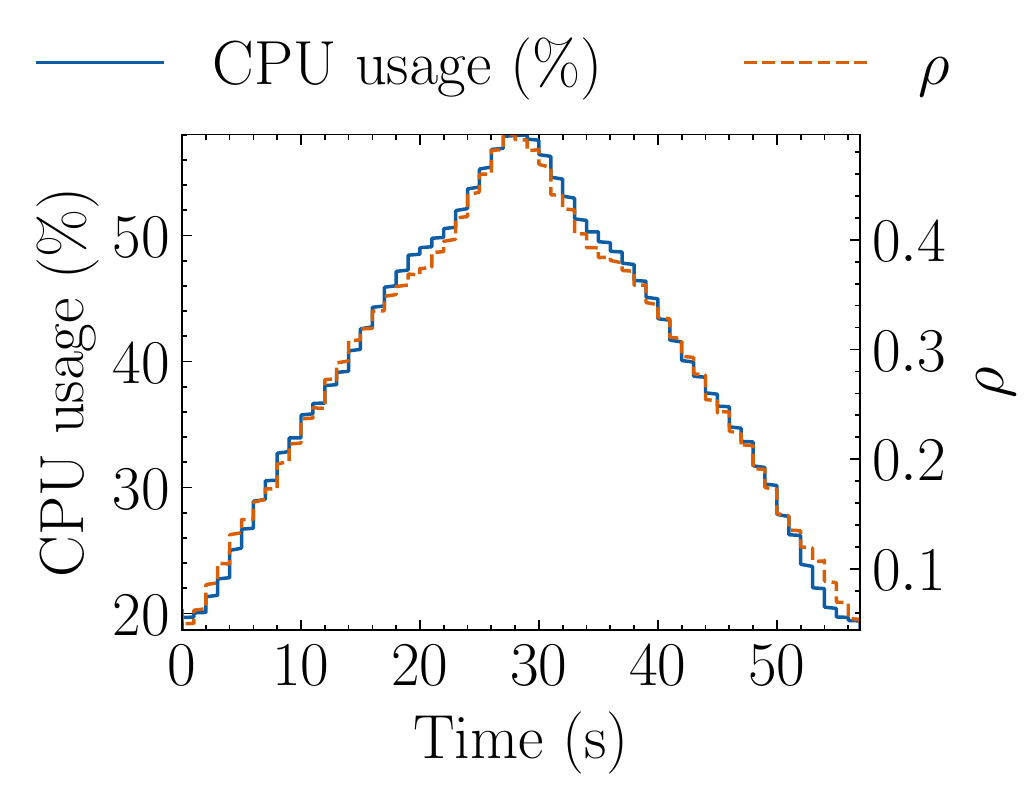}%
        \label{fig:adaptivecpu}%
        }%
    \caption{Metronome's correct adaptation to the incoming traffic load}
    \vspace{-0.4cm}
\end{figure}

To test the dynamic capabilities of Metronome to adapt to varying workloads, we modified the Moongen \texttt{rate-control-methods.lua} example to generate constant bit rate traffic at a variable speed: in a time interval of one minute, Moongen increases the sending rate every 2 seconds until  14 Mpps of rate is reached at about 30 seconds, and then it starts decreasing. Figure \ref{fig:adaptiverate} shows how Metronome \emph{perfectly} matches the Moongen generated traffic rate and how the $T_S$ parameter---set by the threads proportionally---adapts. Figure \ref{fig:adaptivecpu} proves that Metronome promptly adapts CPU usage with respect to the incoming traffic, starting from about 20\% with no traffic and increasing up to 60\% under almost line rate conditions. Also the $\rho$ parameter correctly adjusts its value along with the traffic load.

\subsection{Comparing Metronome and DPDK}\label{sec_comparison}
\begin{figure}[t]
    \centering
    \subfloat[Latency boxplot]{%
        \includegraphics[width=0.5\linewidth]{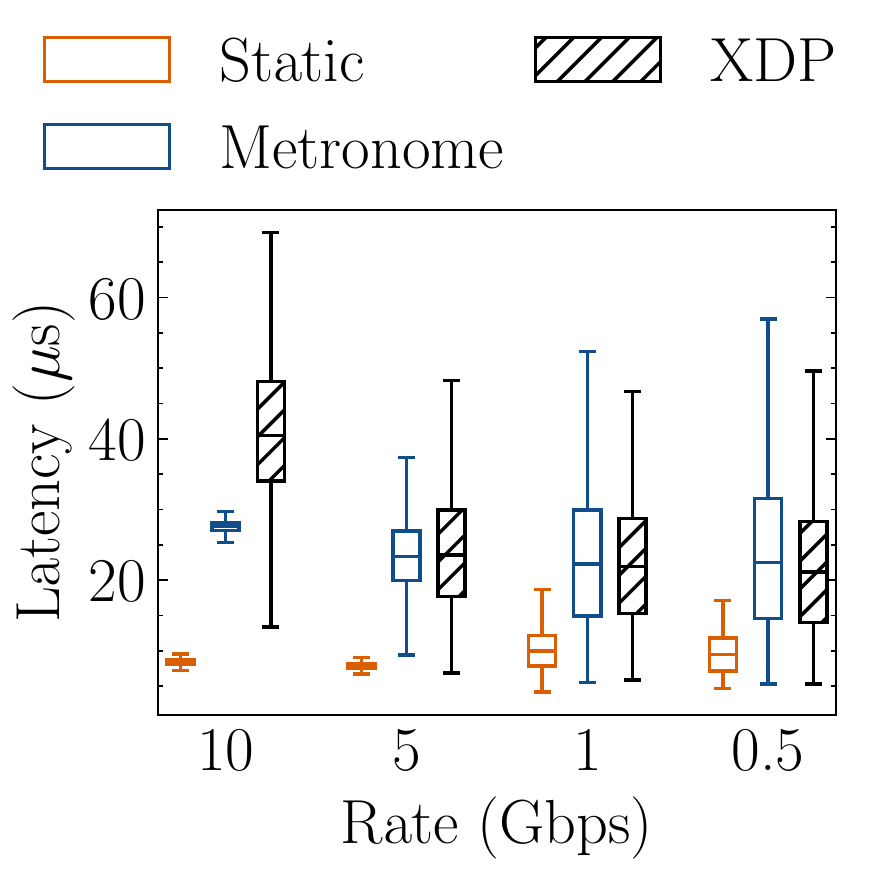}%
        \label{fig:latencyvsdpdk}%
        }%
    \subfloat[CPU usage]{%
        \includegraphics[width=0.52\linewidth]{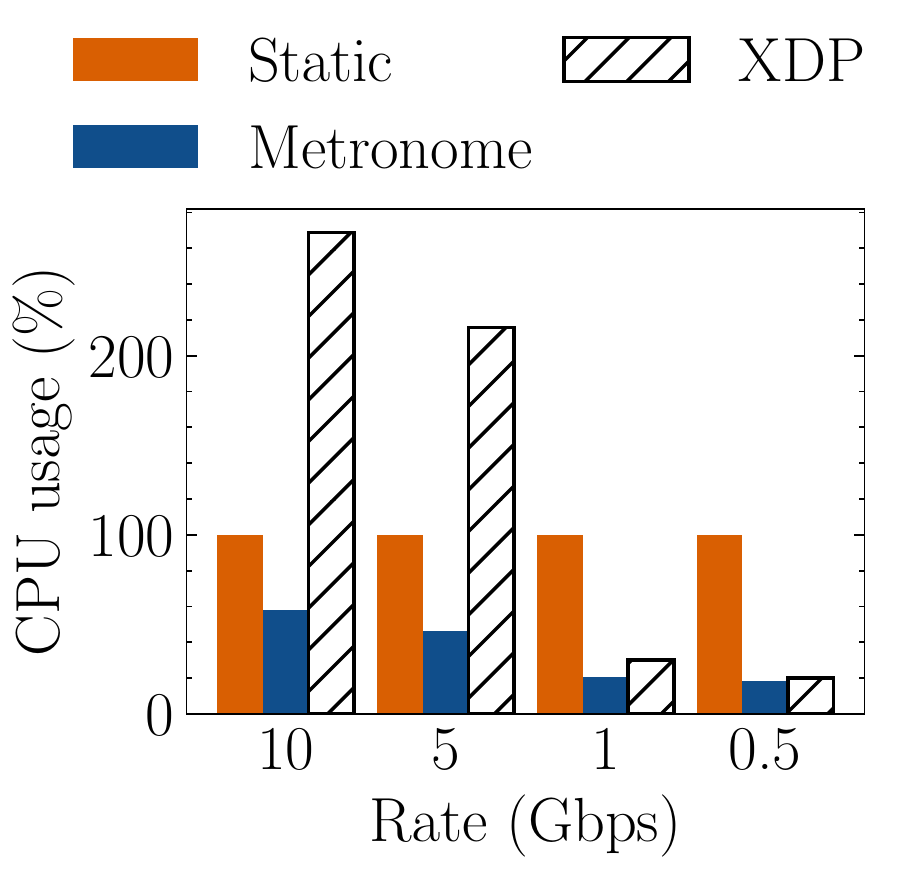}%
        \label{fig:dpdkvsmetronomecpu}%
        }%
    \caption{L3 Forwarder example running static DPDK, Metronome and XDP}\label{fig:l3fwd}
    \vspace{-0.4cm}
\end{figure}

\begin{figure}[t]
\vspace{-0.2cm}
    \centering
    \subfloat[\texttt{ondemand}]{%
        \includegraphics[width=0.5\linewidth]{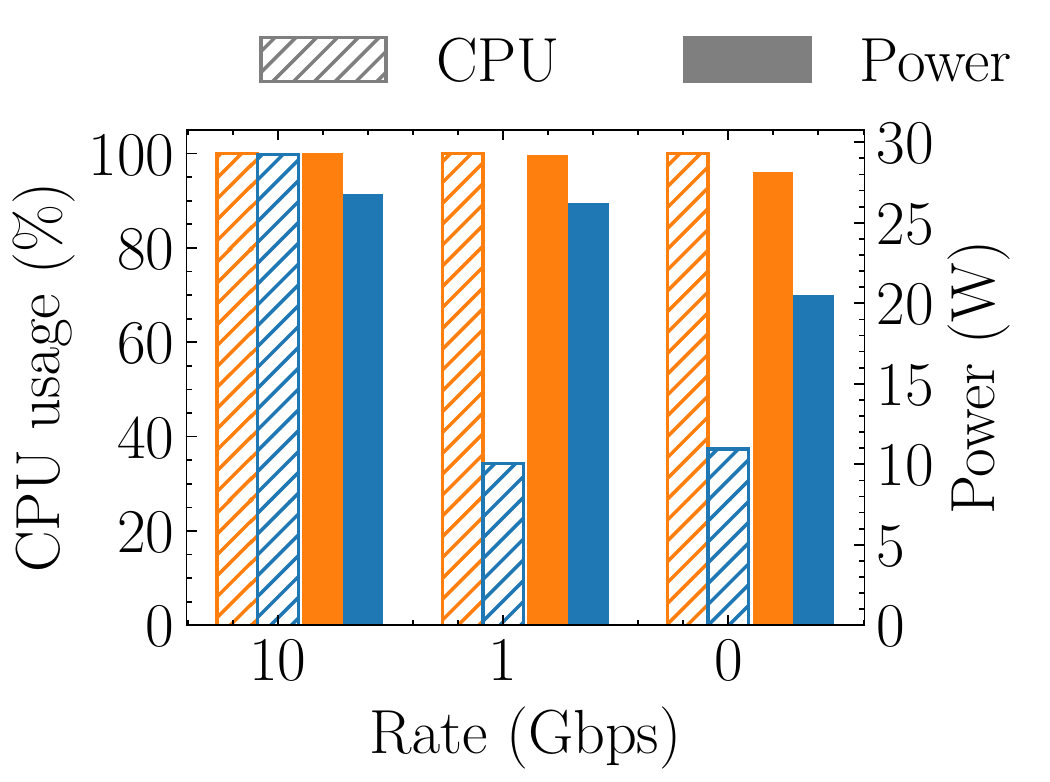}%
        \label{fig:pg-on-demand}%
        }%
    \subfloat[\texttt{performance}]{%
        \includegraphics[width=0.52\linewidth]{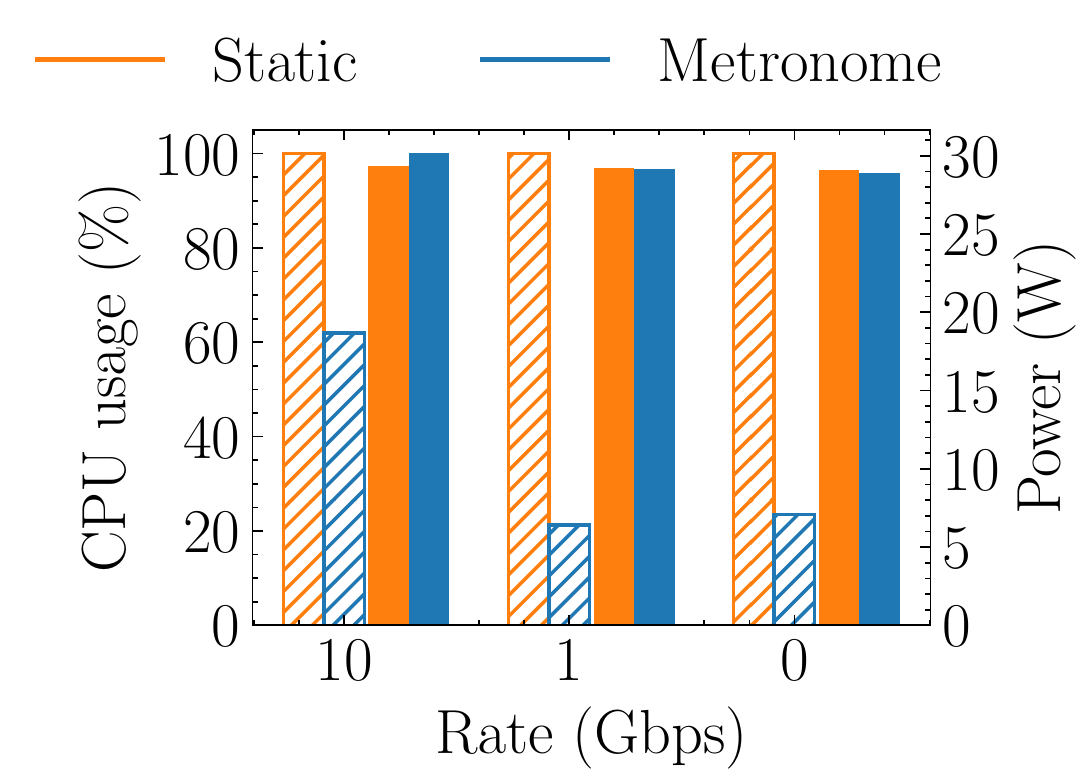}%
        \label{fig:pg-performance}%
        }%
    \caption{Power vs CPU utilization for different power governors.}
    \vspace{-0.6cm}
\end{figure}

We now focus on the comparison between the adaptive Metronome capabilities and the static, continuous polling mode of DPDK in terms of (i) induced latency, (ii) overall CPU usage and (iii) power consumption.\\
\noindent \textbf{Latency}: 
we tested Metronome in order to investigate how the sleep\&wake approach impacts the end-to-end latency. One of our goals was to experiment a constant vacation period, therefore a constant mean latency. Figure \ref{fig:latencyvsdpdk} shows how Metronome (blue boxplots) successfully fulfills this requirement, despite a negligible increase under line-rate conditions, which seems obvious. 
DPDK clearly benefits from its continuous polling operations as it induces about half of the mean latency that Metronome achieves and is also more reliable in terms of variance (see Figure \ref{fig:latencyvsdpdk} - orange boxplots). However, rather than very low latency, Metronome targets an adaptive and fair usage of CPU resources with respect to the actual traffic. 
The minimum latency that Metronome can induce is mainly limited by two aspects: the first one is the Tx batch parameter. Since DPDK transmits packets in a minimum batch number which is tunable, as our system periodically experiments a vacation period some packets may remain in the transmission buffer for a long period of time without actually being sent: this is clearly visible as variance at low rates increases. To overcome such a limitation, we ran another set of tests with the transmission batch set to 1, so that no packets can be left in the Tx buffer. We found out positive impacts on both variance and (slightly) mean values for very low rates. Downgrading the Tx threshold to 1 comes at the cost of a 2-3\% increase in CPU utilization at line rate.
The second aspect is the minimum granularity that \texttt{hr\_sleep()} can support, even if the sleep time requested is much smaller than microseconds (i.e., some nanoseconds). By tuning the first parameter and patching \texttt{hr\_sleep()} in order to immediately return control if a sub-microsecond sleep timeout is requested, we managed to obtain a 7.21 $\mu$s mean delay in Metronome which is very close to the DPDK minimum one (6.83 $\mu$s), and also a significant decrease in variance (0.62 $\mu$s in Metronome vs. 0.43 $\mu$s in DPDK) while still maintaining a 10\% advantage in CPU consumption.\\
\noindent 
\textbf{Total CPU usage}: Figure \ref{fig:dpdkvsmetronomecpu} shows the significant improvements by Metronome (blue bars): while DPDK's greedy approach (orange bars) gives rise to fixed 100\% CPU utilization, Metronome's adaptive approach clearly outperforms DPDK as it is able to provide 40\% CPU saving even under line-rate conditions, while under low rate conditions the gain further rises to more than 5x (Metronome achieves around 18.6\% CPU usage at 0.5Gbps). We underline that Metronome's CPU consumption could be further decreased by increasing the $T_L$ value as explained in Section \ref{sec:tuning}.\\
\noindent \textbf{Power consumption}: as for energy efficiency, it is critical to examine the two approaches depending on the different power governors\cite{powergovernors} available in Linux. More specifically, we concentrated on the two most performing ones, namely \texttt{ondemand} and \texttt{performance}.
The first can operate at the maximum  possible speed, but dynamically adapts the CPU frequency by periodically examining the current CPU load and depending on some threshold values, while the second one keeps the CPU cores at their maximum speed while executing code. While \texttt{ondemand} permits a more adaptive CPU policy, it is less reactive than \texttt{performance}. In particular, CPU cores need more time to get to the maximum speed, but this permits some savings in terms of power.
This trade-off is clearly visible in Figures \ref{fig:pg-on-demand} and \ref{fig:pg-performance}: except for the 10Gbps throughput under the \texttt{performance} power governor scenario, Metronome achieves less power consumption than the traditional DPDK does, with the maximum gain reached when operating under no traffic with the \texttt{ondemand} governor (around 27\%). We underline that in the \texttt{ondemand} scenario Metronome's CPU usage is higher than in the previously seen plots. While we concentrated on the \texttt{performance} governor since we wanted to minimize Metronome's CPU consumption, these tests show that depending on the user/provider's needs, Metronome can also achieve significant power saving when compared to static polling DPDK.

\subsection{Comparing Metronome and XDP}\label{sec:xdp}
We believe it is the case for Metronome to be also compared against XDP \cite{xdp}: this work has a similar motivation to Metronome's main one (reduced, proportional CPU utilization) and is nowadays integrated into the Linux kernel. Despite this similar goal, the approach of the two architectures is quite different: XDP is based on interrupts and every Rx queue in XDP is associated to a different, unique CPU core with a 1:1 binding. Through a conversation with one of the XDP authors on GitHub \cite{xdpissue}, we discovered that our Intel X520 NICs (running the \texttt{ixgbe} driver) achieve at their best a close-to line-rate performance: in fact, the maximum we managed to get is 13.57 Mpps with 64B packets. To do this, we had to equally split flows between four different cores running the \texttt{xdp\_router\_ipv4} example (the most similar one to DPDK's \texttt{l3fwd}). The graphs now discussed are obtained using the minimal number of cores for XDP in order not to lose packets\footnote{We decreased the Mpps sending rate to 13.57 by sending 72B packets, so that XDP wasn't losing packets.} (4 cores on 10Gbps and 5Gbps, 1 core on 1Gbps and 0.5Gbps). We remark that if XDP is deployed with the goal of potentially sustaining line-rate performance, on our test server it should statically be deployed on four cores since there's no way to dynamically increase the number of queues (and therefore, cores) without the user's explicit command through \texttt{ethtool}: in that case, XDP's total CPU usage increases at 52\% @1Gbps and 34\% @0.5Gbps.
Figure \ref{fig:latencyvsdpdk} shows the latency boxplot for XDP: while (even with interrupt mitigation features enabled) we see an increased latency at line rate, we experimented similar latencies at lower rates (we underline that decreasing Metronome's $\bar V$ and the Tx batch parameter we could obtain lower latency results as shown in Section \ref{sec_comparison}, while XDP is already operating at its best performance). Figure \ref{fig:l3fwd} shows XDP's mean total CPU utilization, which is clearly much higher because of the 
per-interrupt housekeeping instructions required to lead control to the packet processing routine, which have an incidence especially at higher packet rates.
On the other hand, XDP occupies no CPU cycles at all under no traffic, while Metronome still periodically checks its Rx queues. This different approach permits Metronome to be highly reactive in case of packet burst arrivals (as shown in Section \ref{sec:tuning}), while XDP loses some tens of thousands of packets in this case before adapting.

\subsection{Impact}\label{sec:impact}
\begin{figure}[]
\vspace{-0.4cm}
    \centering
	\includegraphics[width=0.3\textwidth]{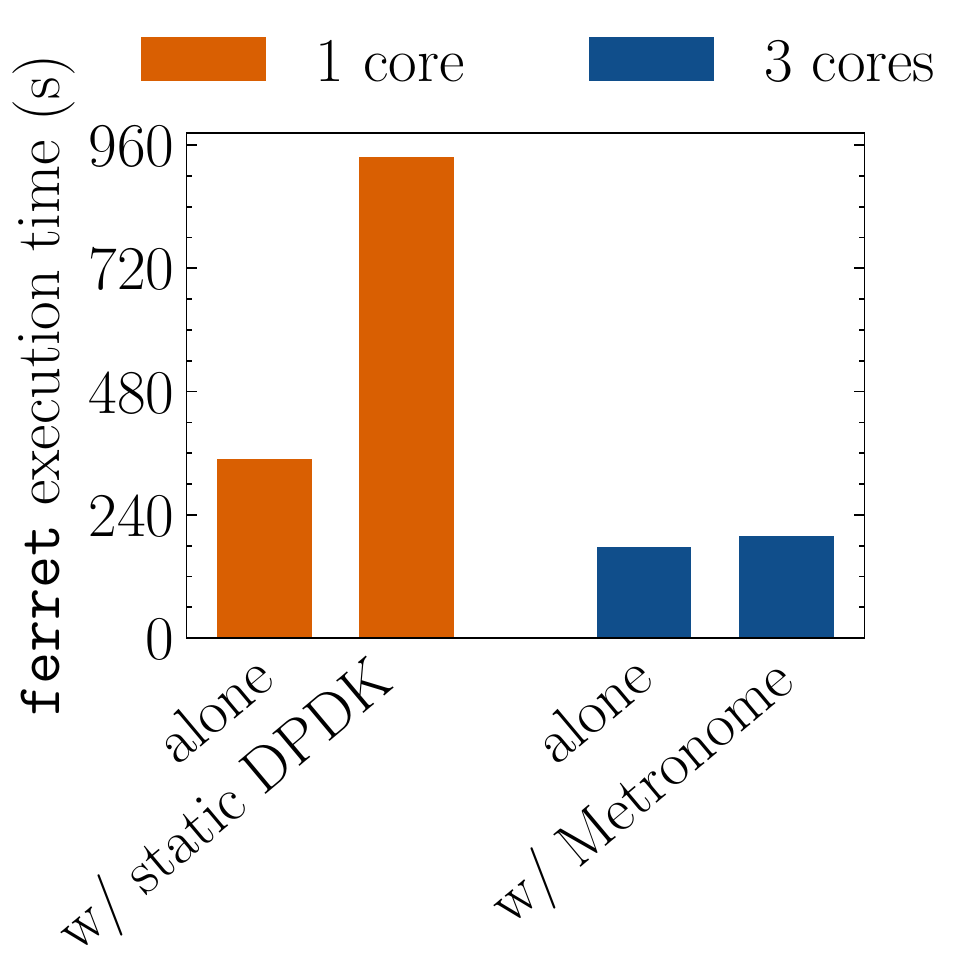}
	\caption{Execution time for \texttt{ferret}}
	\label{fig:ferretexec}
	\vspace{-0.6cm}
\end{figure}

Finally, we analyze Metronome's capabilities to work in a normal CPU sharing scenario, where different tasks compete for the same CPU. We first focus on motivating our multi-threading approach, then we show that the CPU cycles not used by Metronome can be exploited to run other tasks in the meantime without significantly affecting Metronome's capabilities. In both the experiments, Metronome is sharing its same three cores with a VM running \texttt{ferret}, a CPU-intensive, image similarity search task coming from the PARSEC \cite{parsec} benchmarking suite.
Because the Metronome task is more time sensitive than the \texttt{ferret} one, we give Metronome a slight scheduling advantage by setting its niceness value to -20, while the VM's niceness is set to 19 since it has no particular time requirements. In any case, the two are still set to belong to the same {\tt SCHED\_OTHER} (normal) priority class.\\
\noindent \textbf{The case for multiple threads}: While we previously stated that a few threads are better for Metronome, we now clarify the reason for using multiple threads by scheduling the VM running the \texttt{ferret} program on one core. When running Metronome on the same single core, because of the CPU conflicting scenario the maximum throughput achievable by \texttt{l3fwd} is around 8 Mpps. If we deploy Metronome on three cores (one of these three cores is the same used by the VM), only one thread will be highly impacted by the CPU-intensive task and therefore will unlikely act like a primary thread. In this case \texttt{l3fwd} achieves no packet loss on a 10Gbps link, and the same scenario happens if we schedule the same VM running \texttt{ferret} on two of the three cores shared with Metronome. The next paragraph shows that also when all of the three Metronome threads are (potentially) impacted by \texttt{ferret}, they can still forward packets at line rate, thanks to the reduced likelihood that all of them (when requiring to be brought back to the runqueue after the sleep period) are impacted simultaneously because of the decisions of the OS CPU-scheduler. These experiments clearly show that running Metronome on multiple threads leads to improved robustness against common CPU sharing scenarios and interference by other workloads.\\
\noindent \textbf{Co-existence with other tasks}: we now demonstrate that Metronome's sleep\&wake approach enables the CPU sharing of other tasks without major drawbacks, while DPDK's static, constant polling approach denies such possibilities. We first ran \texttt{ferret} on one core, 
with a static DPDK polling \texttt{l3fwd} application on the same core. Then, we scheduled \texttt{ferret} on three cores and  the three Metronome threads on the same cores. As Figure \ref{fig:ferretexec} shows, sharing the CPU with a static polling task causes \texttt{ferret} to almost triple its duration, while Metronome's multi-threading and CPU sharing approach only causes a 10\% increase. Moreover, standard DPDK's single core approach couldn't keep up with the incoming load, achieving a maximum of 7.31 Mpps, while Metronome achieved no packet loss even when all of its three cores were shared with a CPU intensive program such as \texttt{ferret} (see Table 4). We underline that Metronome's multi-threading strategy implies that the same workload is shared between multiple threads, thus the more the cores, the less the work every thread needs to perform and therefore the more they can co-exist with other tasks without affecting performances, as this test shows.

\begin{table}[]
\centering
\begin{tabular}[]{|c|cc|}\hline
            & alone & w/ \texttt{ferret} \\ \hline
            static DPDK & 14.88 & 7.34 \\
            Metronome & 14.88 & 14.88 \\ \hline
     \end{tabular}\label{tab:ferret_pktloss}
    \captionof{table}{Throughput (Mpps) for static DPDK and Metronome}
\end{table}

\subsection{Going multiqueue}\label{sec:eval_multi}
Our evaluation now focuses on the multiqueue case analyzed in Section \ref{sec:multi}: tests have been conducted for both Metronome and static DPDK using Intel XL710 40Gbps NICs. These devices are limited by a maximum processing rate of 37Mpps \cite{intelxl710}. In all tested environments, Metronome always reached the desired 37Mpps forwarding throughput. Traffic is distributed equally among the RX queues through RSS, while in a later subsection we will discuss the unbalanced traffic case. We found out that the main components to be tuned for achieving the best performances in Metronome (assuming a fixed ${\bar V}=15\mu s$) are the number of Rx queues and the CPU power governor.
\subsubsection{Tuning the number of queues}  We test our l3fwd application using 2,3 and 4 Rx queues for the same 37Mpps throughput. Results for CPU and power consumption are available in Figure \ref{fig:multiqueue_cpupower}. We now focus on the \texttt{performance} power governor (Figures \ref{fig:multiqueue_cpupower}a,b,c), as we discuss the impact of the \texttt{ondemand} power governor in the next paragraph. The $\rho$ parameter and the busy tries percentage are also shown (see Figure \ref{fig:multiqueue_rhobusy}) in order to better explain the results.
As with 2 Rx queues every queue is experiencing high load traffic (\textasciitilde18Mpps each), most of the time the queues are busy ($\rho=0.7$ with 2 threads) and the CPUs are running at their maximum, so the main gain is in the CPU occupancy (150\% with 2 threads, ~156\% with 8). While in the cases with many threads Metronome uses more power than static DPDK (here represented with dotted lines), it does not make much sense to use more than 4 threads to contend just two queues, as also the linear increase in busy tries (blue-filled bars in Figure \ref{fig:multiqueue_rhobusy}a) suggests.
When using a larger number of queues (3 or 4), the lower per-queue load permits Metronome to increase its gain compared to static DPDK both in CPU and power (see Figure \ref{fig:multiqueue_cpupower}c). It is also worth noticing that with a larger number of queues, $\rho$ also decreases and, consequently, the number of busy tries decreases, which makes the Metronome algorithm more efficient.
\subsubsection{Power governors matter} 
\begin{figure*}[]
\centering
\begin{minipage}{0.31\textwidth}
  \centering
\includegraphics[width=\textwidth]{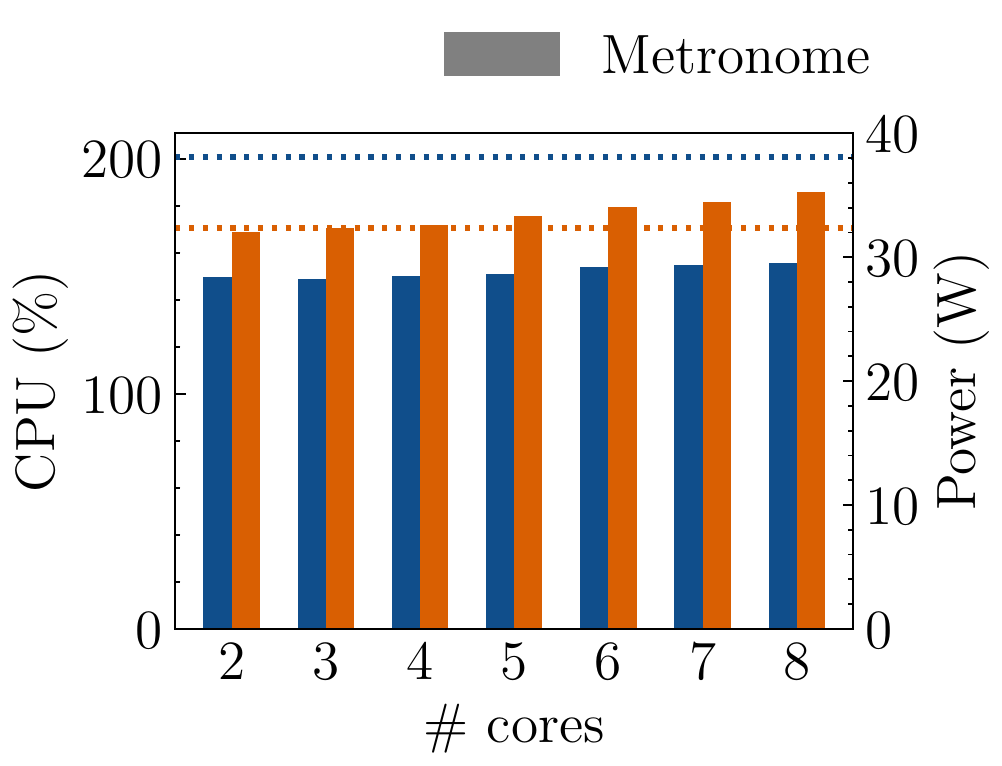}
\subcaption{2 queues - \texttt{performance}}
\end{minipage}%
\begin{minipage}{0.31\textwidth}
  \centering
\includegraphics[width=\textwidth]{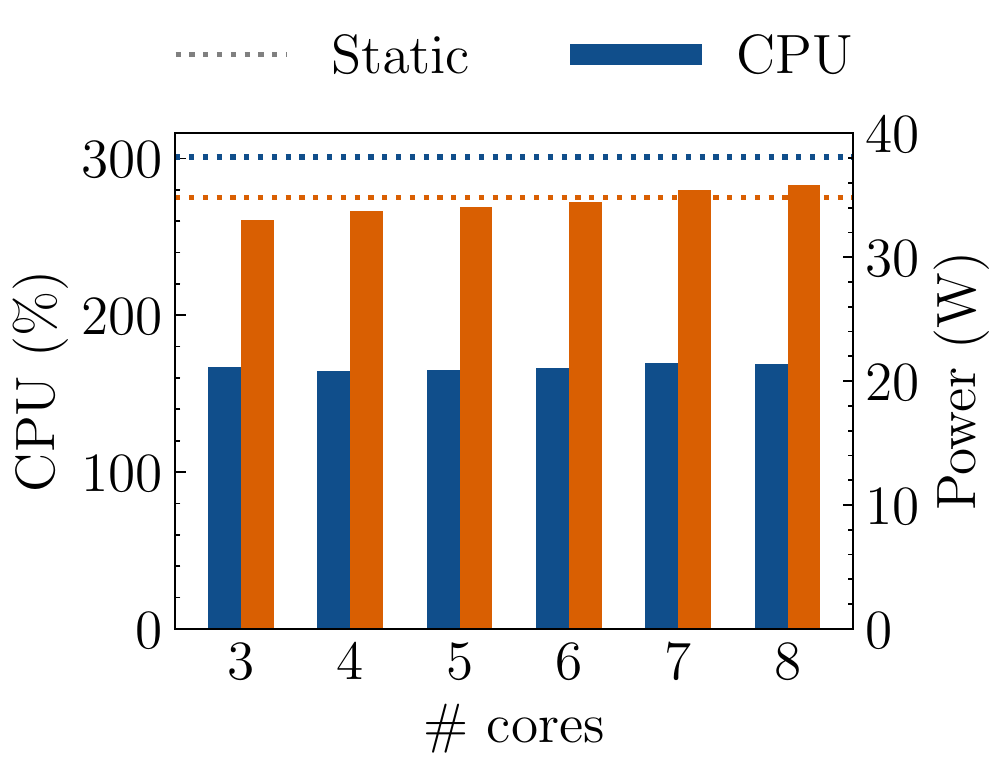}
\subcaption{3 queues - \texttt{performance}}
\end{minipage}%
\begin{minipage}{0.31\textwidth}
  \centering
\includegraphics[width=\textwidth]{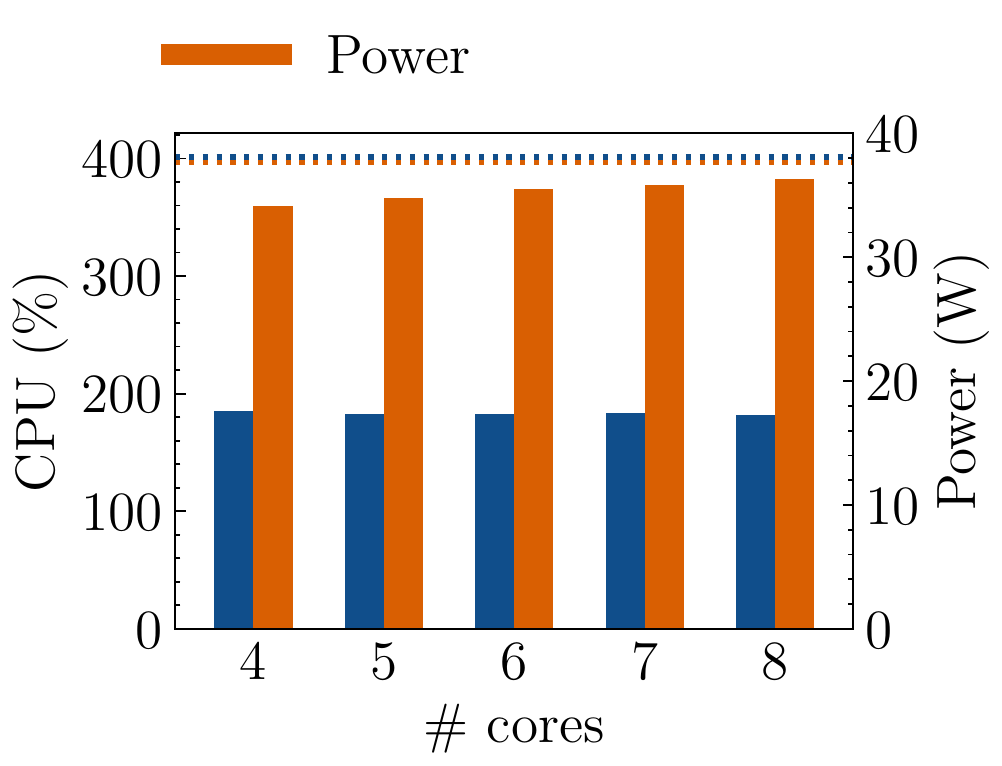}
\subcaption{4 queues - \texttt{performance}}
\end{minipage}

\begin{minipage}{0.31\textwidth}
  \centering
\includegraphics[width=\textwidth]{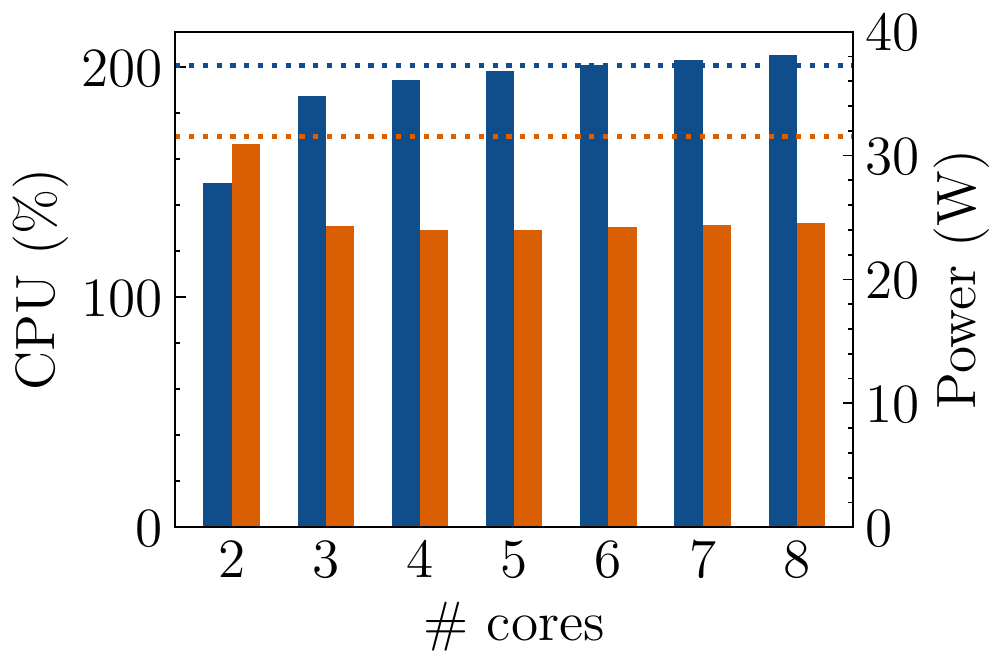}
\subcaption{2 queues - \texttt{ondemand}}
\end{minipage}%
\begin{minipage}{0.31\textwidth}
  \centering
\includegraphics[width=\textwidth]{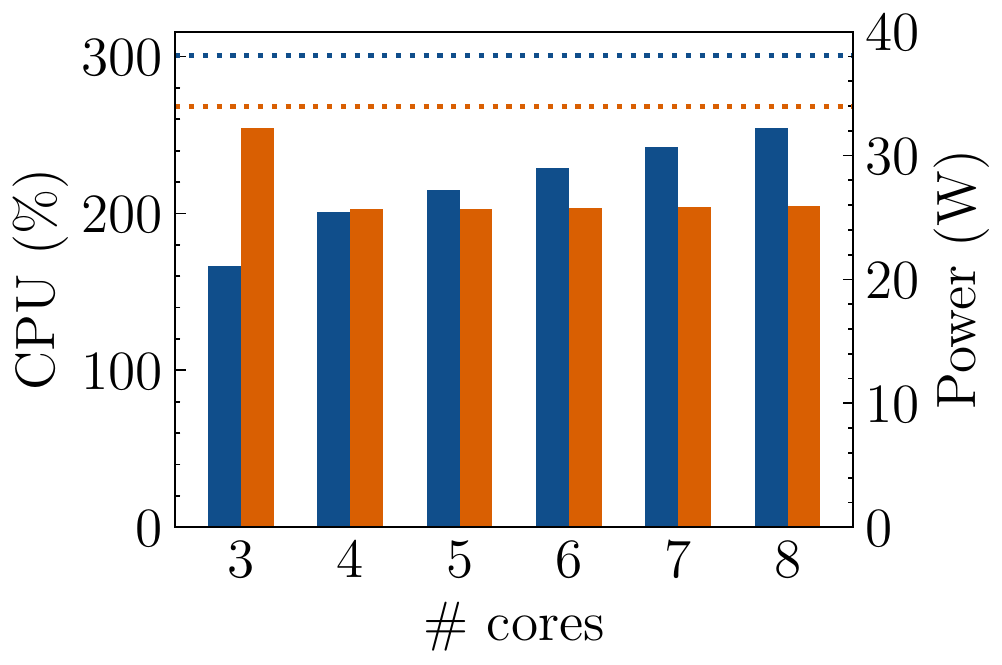}
\subcaption{3 queues - \texttt{ondemand}}
\end{minipage}%
\begin{minipage}{0.31\textwidth}
  \centering
\includegraphics[width=\textwidth]{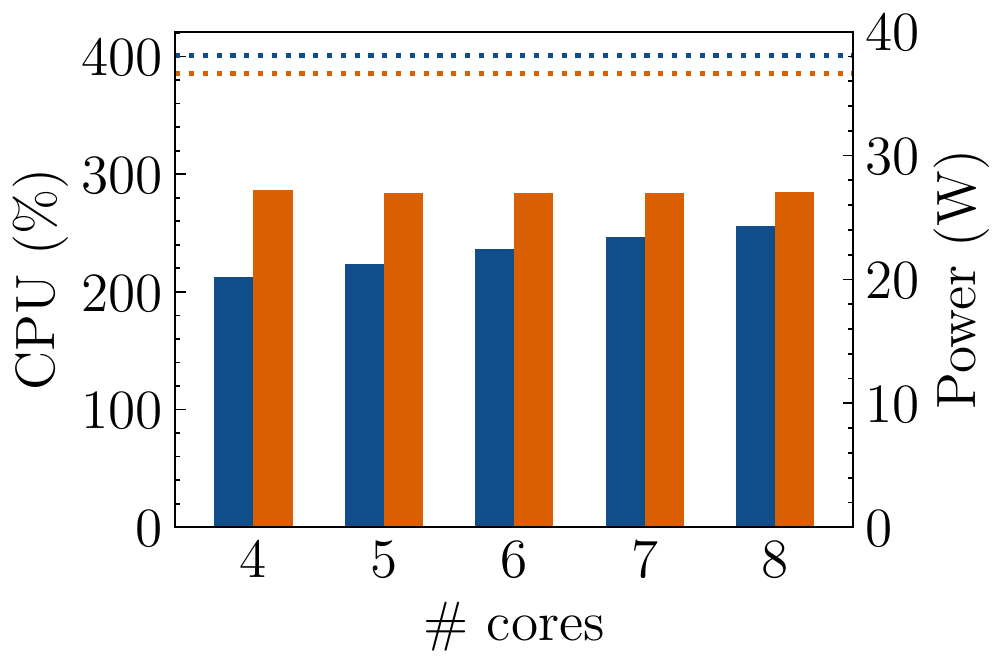}
\subcaption{4 queues - \texttt{ondemand}}
\end{minipage}
\caption{CPU and power for Metronome and static DPDK with different RX queues and power governors} \label{fig:multiqueue_cpupower}
\end{figure*}

\begin{figure*}[]
\centering
\begin{minipage}{0.31\textwidth}
  \centering
\includegraphics[width=\textwidth]{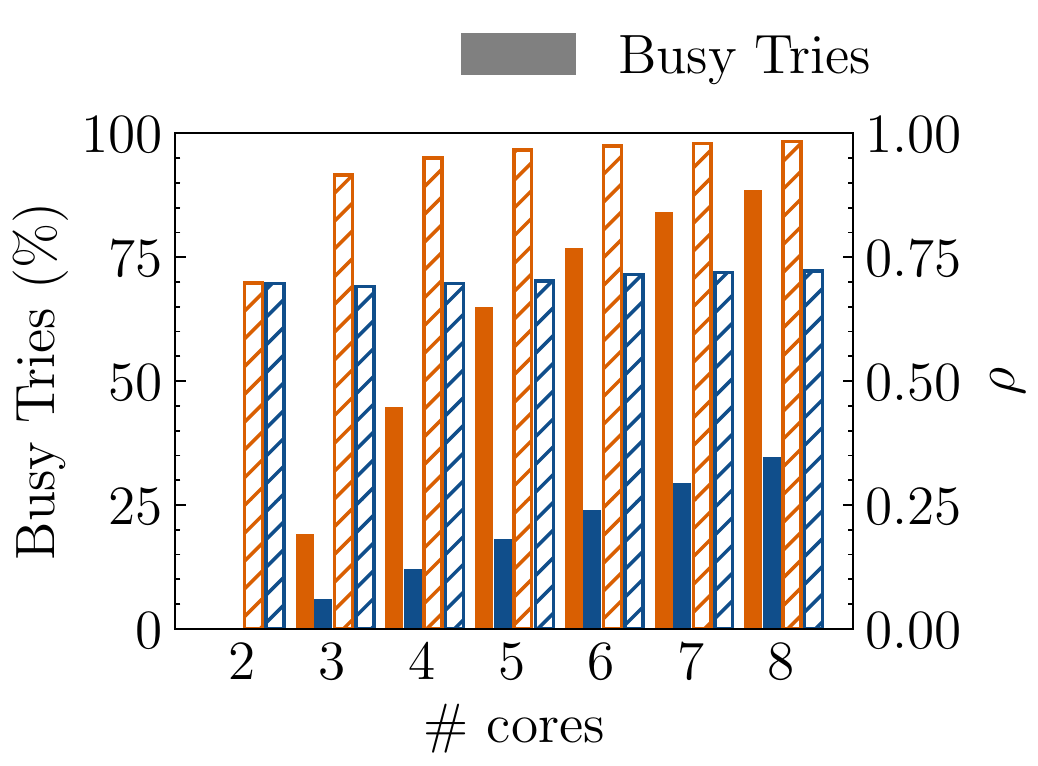}
\subcaption{2 queues}
\end{minipage}%
\begin{minipage}{0.31\textwidth}
  \centering
\includegraphics[width=\textwidth]{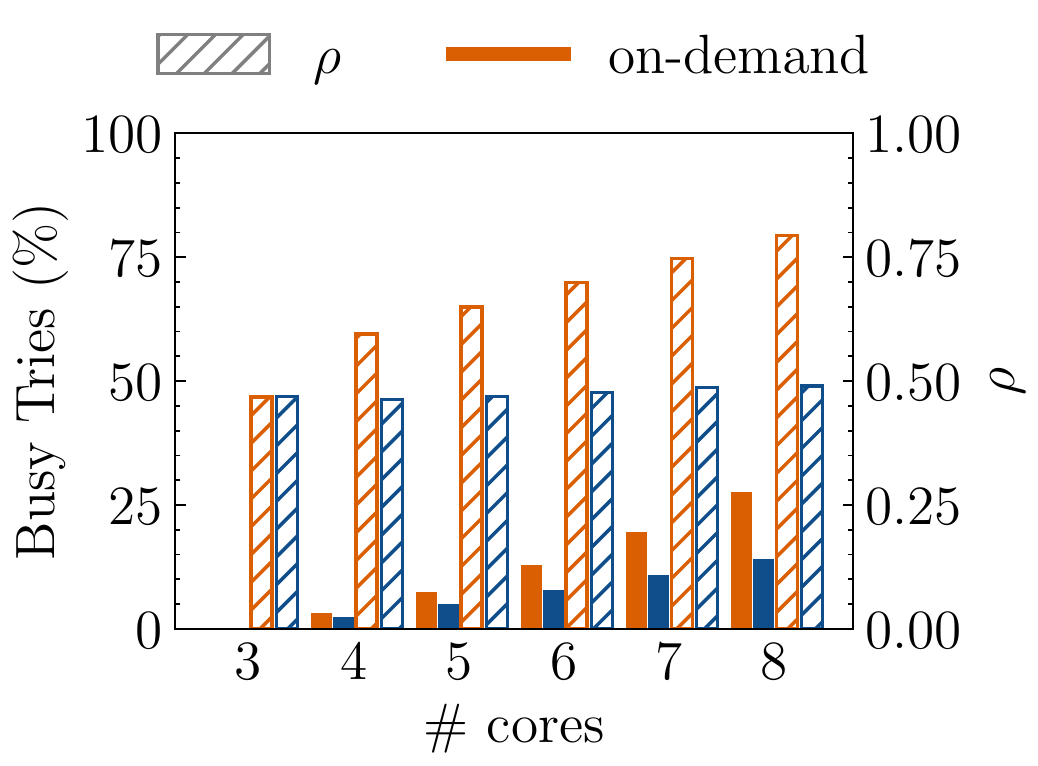}
\subcaption{3 queues}
\end{minipage}%
\begin{minipage}{0.31\textwidth}
  \centering
\includegraphics[width=\textwidth]{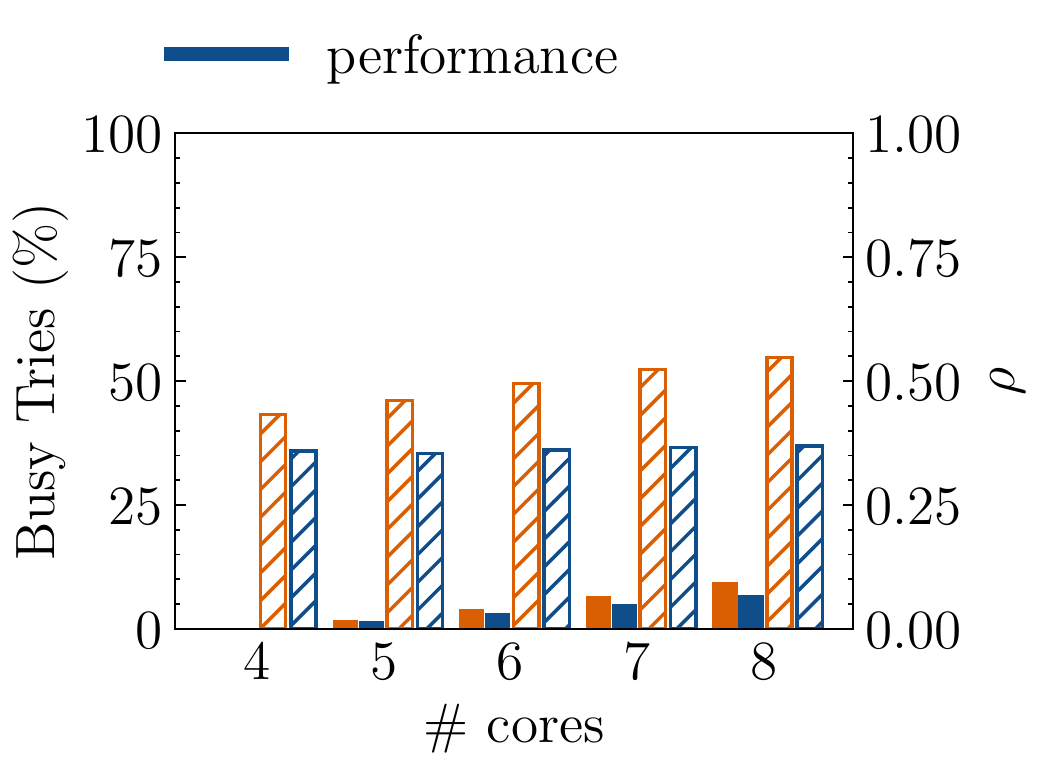}
\subcaption{4 queues}
\end{minipage}
\caption{Busy tries and $\rho$ with different \# of Rx queues} \label{fig:multiqueue_rhobusy}
\end{figure*}

While in the previous paragraph we focused on the \texttt{performance} governor, we now discuss the \texttt{ondemand} one, the difference between the two is explained in Section \ref{sec_comparison}.
Figure \ref{fig:multiqueue_cpupower}d shows the results with 2 Rx queues: the initial decrease in power consumption is motivated by the fact that while with 2 threads, these can only be in the primary state, when increasing the number of threads, they tend to be backup ones (and therefore, to sleep for more time) because of the high percentage of busy tries (see the red-filled bars in Figure \ref{fig:multiqueue_rhobusy}a). This is in turn caused by the steep increase of $\rho$ with the number of threads: since the CPU cores can execute at slower rates, threads will likely take more time to unload their Rx queues and therefore these will be busy for longer periods. This phenomenon is still visible with 3 queues and slightly with 4 queues. As the number of queues increases, the difference between the two power governors in terms of queue occupation $\rho$ and busy tries still remains significant but also slightly decreases (see the subfigures in Figure \ref{fig:multiqueue_rhobusy}). Overall, the \texttt{ondemand} power governor permits to trade some extra CPU time for a better power efficiency: also in this case the best advantages are visible with a larger number of queues. This further demonstrates Metronome's capability to adapt to a lower per-queue load.

\subsubsection{Scaling to the actual traffic}
\begin{figure}[]
    \centering
    \subfloat[CPU usage]{%
        \includegraphics[width=0.5\linewidth]{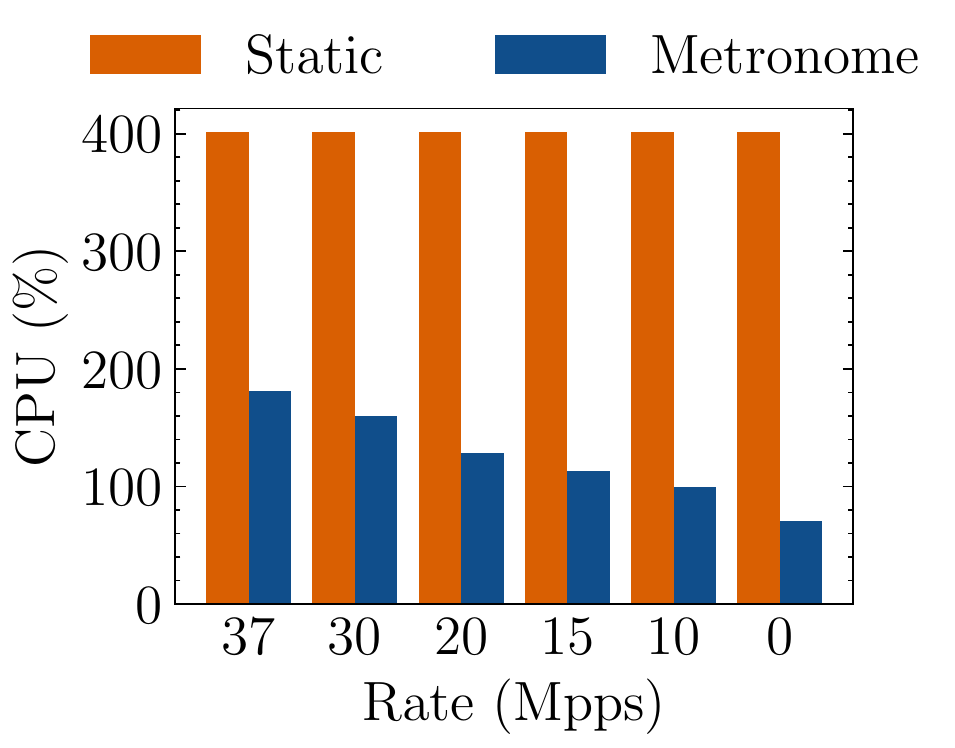}%
        \label{fig:mq_cpu}%
        }%
    \subfloat[Power]{%
        \includegraphics[width=0.48\linewidth]{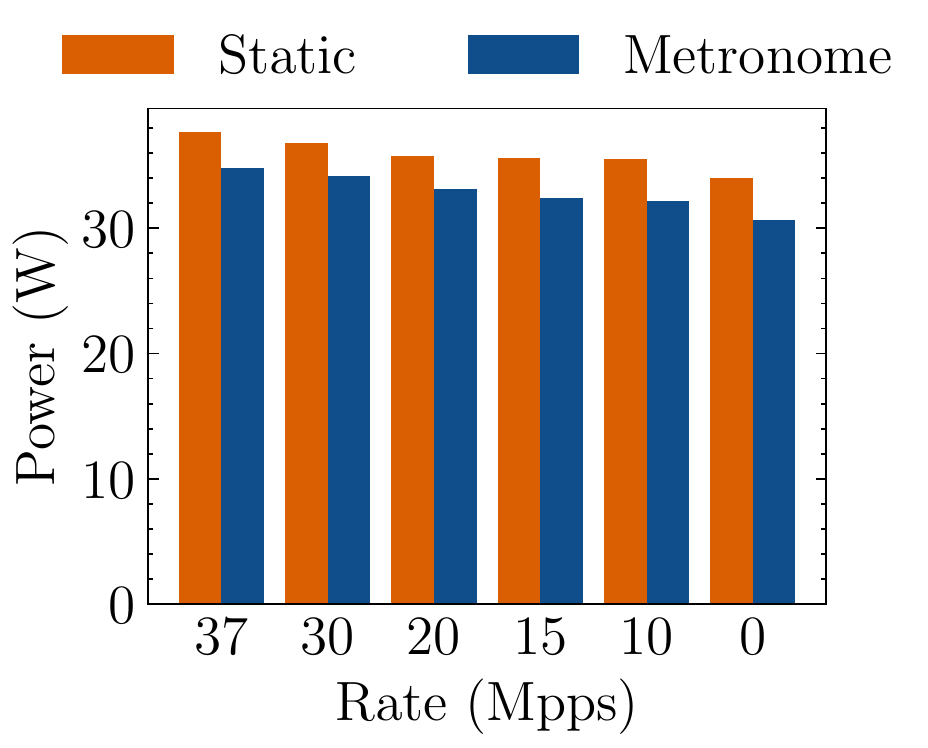}%
        \label{fig:mq_power}%
        }%
    \caption{CPU and power consumption under different loads. The power governor is \texttt{performance}}
   \vspace{-0.4cm}
\end{figure}

Figure \ref{fig:mq_cpu} shows the CPU consumption for Metronome and DPDK under different traffic rates, from 0 to line rate on an Intel XL710 (37Mpps). The test is done with 4 Rx queues
with both Metronome and DPDK, and with $M=5$ and $\bar V=15\mu s$ for Metronome. Our approach saves more than half of static DPDK's CPU cycles while maintaining the same line-rate throughput, and improving even more at lower rates. Also in terms of power consumption (see Figure \ref{fig:mq_power}), Metronome provides around 2-3W of advantage even when using a highly expensive power governor such as \texttt{performance}.
\subsubsection{Unbalanced traffic}
We test Metronome's multiqueue capabilities by continuously sending at line rate an unbalanced pcap file. The file is composed by 1000 packets, 30\% of the packets belongs to the same UDP flow, while the other 70\% is randomly generated and therefore equally split among the queues. In the test we use 3 Rx queues (without losing packets), so the most stressed queue processes around 53\% of the total throughput, while the other two queues are in charge of 23\% each. Table \ref{tab:mq} shows some meaningful statistics for the test. Queue \#2 is the most stressed. Therefore, it has the highest busy tries percentage and also the highest queue occupation $\rho$. It is worth noticing that, on a 3-minute test, queue \#2 experienced less than half of the lock tries of queues \#1 and \#3: this trend validates the assumption in Section \ref{s4:metro}, where a busy queue tends to have only one primary thread at a time while a less occupied one is more likely to have more threads in the primary state simultaneously, and therefore, more tries.

\begin{table}[]
\centering
\begin{tabular}{|l|ccc|}
\hline
& Busy tries (\%) & Total tries & $\rho$    \\
\hline
Queue \#1 & 1.94            & 5970660     & 0.3208 \\
Queue \#2 & 4.39            & 2625007     & 0.7269 \\
Queue \#3 & 2.02            & 5704167     & 0.3552 \\ \hline
\end{tabular}
\caption{Statistics for the unbalanced traffic case} \label{tab:mq}
\end{table}

\subsection{Tested applications}\label{sec:testedapps}
\begin{figure}[t]
    \centering
    \subfloat[IPsec Security Gateway]{%
        \includegraphics[width=0.5\linewidth]{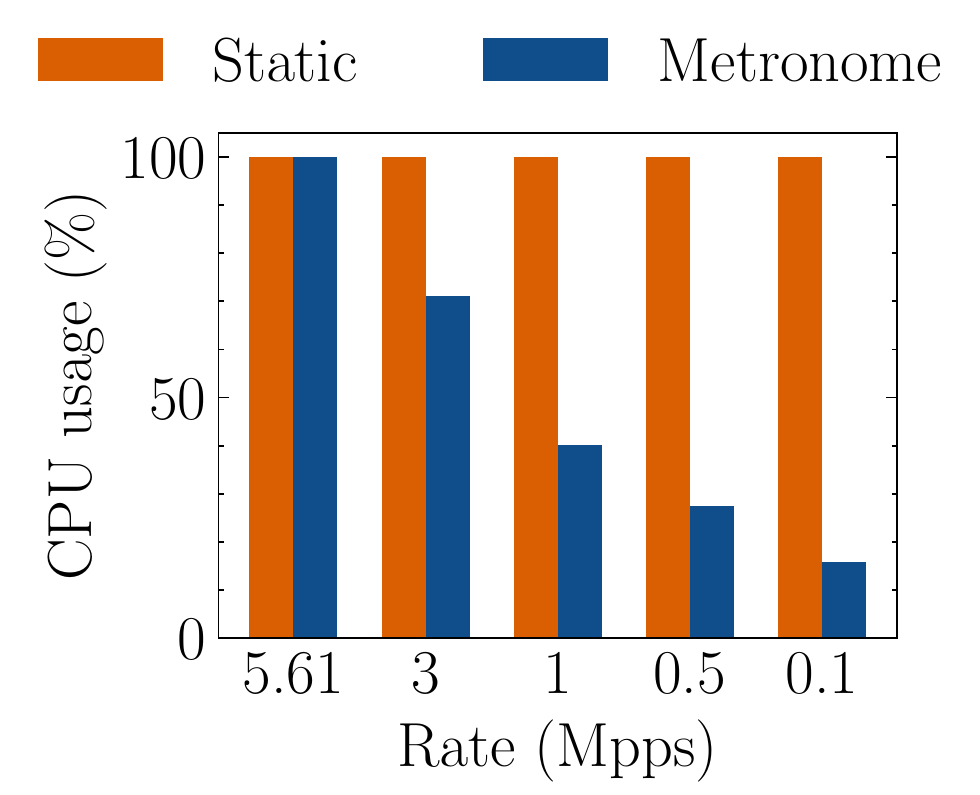}%
        \label{fig:ipsec_cpu}%
        }%
    \subfloat[FloWatcher-DPDK]{%
        \includegraphics[width=0.5\linewidth]{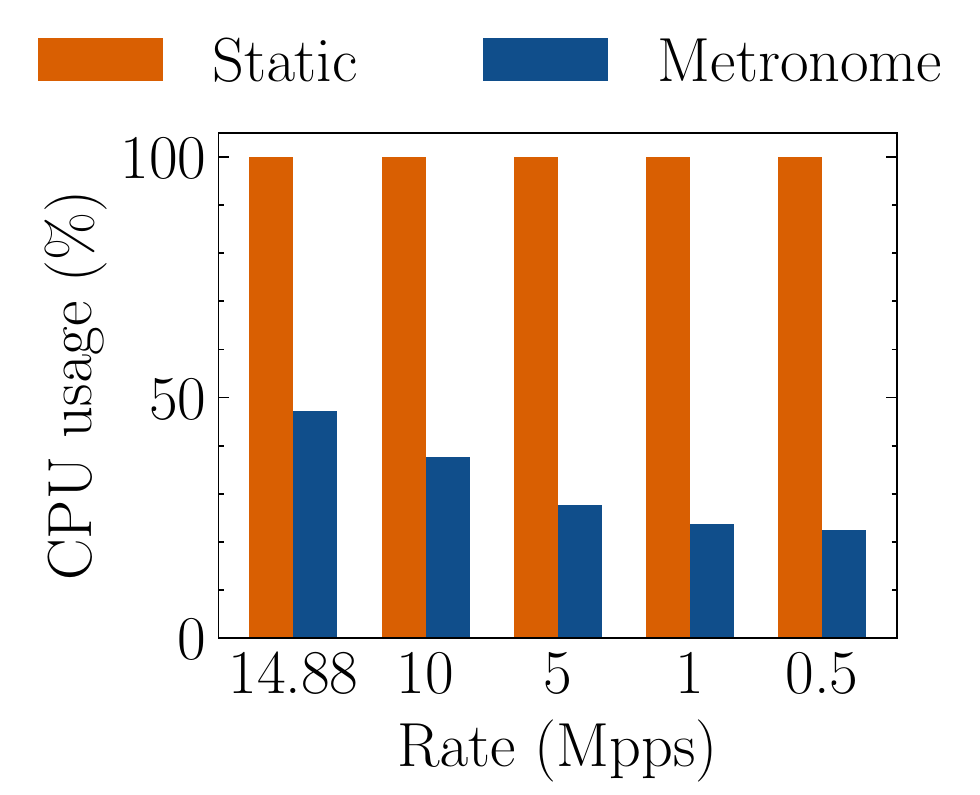}%
        \label{fig:flowmon_cpu}%
        }%
    \caption{CPU usage - other applications (single RX queue)}\label{fig:cpu:ipsec_flowatch}
    \vspace{-0.6cm}
    \end{figure}

To further assess the flexibility and the wide breadth of Metronome, we show three DPDK applications that we successfully adapted to the Metronome architecture, namely two DPDK sample applications as a L3 forwarder \cite{l3fwd} and an IPsec Security Gateway \cite{ipsecgw}, as well as FloWatcher-DPDK \cite{flowatcher}, a high-speed software traffic monitor.\\
\textbf{L3 forwarder}
The \texttt{l3fwd} sample application acts as a software L3 forwarder either through the longest prefix matching (LPM) mechanism or the exact match (EM) one. We chose the LPM approach as it is the most computation-expensive one between the two. We have used the \texttt{l3fwd} application to exhaustively test Metronome's performances in Section \ref{sec:eval}, so we refer the readers to that Section for further performance implications.\\
\textbf{IPsec Security Gateway}
This application acts as an IPsec end tunnel for both inbound and outbound network traffic. It takes advantage of the NIC offloading capabilities for cryptographic operations, while encapsulation and decapsulation are performed by the application itself. Our tests perform encryption of the incoming packets through the AES-CBC 128-bit algorithm as packets are later sent to the unprotected port. The DPDK sample application achieves a maximum outbound throughput of 5.61 Mpps with 64B packets in static polling mode: once we adapted the application to Metronome, we found out that we were able to reach the exact same throughput. In fact, one of Metronome's threads was always processing packets and therefore never releasing the trylock shared with the other threads, this is clearly visible in Figure \ref{fig:ipsec_cpu}. For lower rates, Metronome clearly outperforms the static approach as rates get decreased. 
    
    \noindent \textbf{FloWatcher-DPDK}
    FloWatcher is a DPDK-based traffic monitor application providing tunable and fine-grained statistics, both at packet and per-flow level. FloWatcher can either act through a run to completion model or a pipeline one: we chose the former since the receiving thread is also calculating the statistics, therefore providing a more challenging scenario for Metronome. We find out that Metronome provides the same performances that the static DPDK approach does in terms of zero packet loss and correct statistics calculation, while reaching major improvements in CPU utilization. In particular,  Figure \ref{fig:flowmon_cpu} shows a 50\% gain even under line rate traffic and almost a 5x gain with 0.5 Mpps traffic.

\section{Conclusions}
This paper has proposed and assessed Metronome, an approach devised to replace the continuous and CPU-consuming DPDK polling with a sleep\&wake, load-adaptive, intermittent packet retrieval mode. 
Metronome's viability has been evaluated by integrating it into three different common DPDK applications, and by showing its significant improvements primarily in terms of CPU utilization (and, partially, also in terms of power consumption), and therefore its ability to release precious CPU cycles to business applications. We finally stress that such gains are traded off with an extra latency toll, which can be taken into account and configured using the tuning knobs provided by our approach, especially when (and if) considering the usage of Metronome with time-critical applications.\\
\noindent \textbf{Acknowledgments:} We thank Giuseppe Siracusano and Sebastiano Miano for helping us in tuning XDP.

\bibliographystyle{IEEEtran}
\bibliography{biblio}
\section*{APPENDIX}
\section*{I. Skeleton code} 
\label{app:code}
We briefly compare the classical and Metronome  methods through a simplified (yet meaningful) code example of a typical DPDK thread routine. This example only focuses on the different coding approaches, rather than other aspects (e.g., implementing the actual network functionalities, calculating the optimal timer through our analytical model...). Both examples show a typical packet processing task. The usual DPDK implementation is shown in Listing \ref{lst:dpdk_standard}, while our novel proposal is depicted in Listing \ref{lst:dpdk_novel}.
While both solutions include a set of Rx queues (\texttt{queue[]}) to be processed, in Listing \ref{lst:dpdk_standard} each thread has assigned a specific queue in an exclusive way (line 1), while in Listing \ref{lst:dpdk_novel} queues are shared among multiple threads and therefore require access through the \texttt{trylock()} mechanism (see line 4).
In Listing \ref{lst:dpdk_standard} the thread tries to retrieve a burst of packets (line 4) of maximum size \texttt{BURST\_SIZE}, processes it (line 7) and \textbf{immediately} scans again its set of queues, regardless of the fact that those queues may be experiencing low traffic (or no traffic at all). We highlight that this behavior is the real cause of the 100\% constant CPU utilization by a single thread, as threads are working in a \emph{traffic-unaware} manner. As for the later point, this level of CPU usage is negatively reflected on energy consumption and also on turbo-boost waste. \\
Listing \ref{lst:dpdk_novel} shows our novel approach: once the lock for a certain queue is acquired, the thread processes that queue until it becomes empty (\texttt{while()} loop in lines 10-11), then it releases the lock (line 12) and goes to sleep for a \texttt{timeout\_short} period.
If a certain lock can't be granted, that queue is skipped as a different thread is already processing it: the thread changes its \texttt{curr\_queue}, extracting it randomly from the set of all available queues (line 5), and goes to sleep for a \texttt{timeout\_long} period.\\
Despite the simplicity of these examples, we believe they clearly point out the difference between a \emph{traffic-aware} policy and a static one simply based on greedy resource usage.
\begin{lstlisting}[style=example, caption={A standard DPDK polling loop}, label={lst:dpdk_standard}]
curr_queue = THREAD_ASSIGNED_QUEUE;

while (1) {
  nb_rx = receive_burst(queue[curr_queue], pkts, BURST_SIZE);
  if (nb_rx == 0)
    continue;
  process_and_send_pkts(pkts, nb_rx);
}
\end{lstlisting}
\begin{lstlisting}[style=example, caption={Our novel DPDK processing loop}, label={lst:dpdk_novel}]
curr_queue = THREAD_ASSIGNED_QUEUE;

while (1) { 
  if(!trylock(lock[curr_queue])) {
    curr_queue = randint(n_queues);
    hr_sleep(timeout_long);
    continue;
  }

  while(nb_rx = receive_burst(queue[curr_queue], pkts, BURST_SIZE))
    process_and_send_pkts(pkts, nb_rx);
  unlock(lock[i]);
  
  hr_sleep(timeout_short);
}
\end{lstlisting}
\section{II. Caveats and Details}
\label{s4:caveat:mu}
In this Appendix we discuss some supplementary technical details at the basis of our assumptions. We specifically start from the assumption used in the model presented in Section IV-B: packet retrieval rate $\mu$ independent on the packet size. Even if not strictly necessary\footnote{The renewal arguments brought about in this work remain valid if we replace deterministic quantities with their mean - in other words even if we consider the alternative model of constant retrieval rate in terms of a constant rate of $C$ {\em bits per second}, opposed to $\mu$ {\em packets per second}, we would just need to set $\mu = E[P]/C$, with $E[P]$ being the average packet size.}, our assumption of $\mu$ constant and independent of the packet size is actually motivated by the specific way in which DPDK handles packets. Indeed, DPDK does not process packets by physically moving them from the NIC, but it just moves the relevant descriptors which populate the Rx queue.

Since a typical DPDK application consists of a loop where a receive function is executed at each iteration, the service rate can somewhat be influenced by: (i) the loop length (that is, how much time passes between two consecutive receive operations) and, (ii) how many descriptors are processed by the receive function in each cycle. DPDK usually processes descriptors in a batch defining the maximum number of packets to be processed at each invocation. Usually, this value is set to 32 as it provides a nice tradeoff for the batching benefits without affecting latency. Some interference on the  rate $\mu$ may also be inducted by OS interrupts or because of preemption of DPDK threads by some higher priority thread (like an OS kernel demon). 
However, the multi-tread approach taken by Metronome is devised just to make DPDK more resilient towards this kind of interference scenario, and with no need for dedicated resources---as said, one of our targets is to make DPDK effective in CPU-sharing contexts. In fact, nowadays OS kernels (like the Linux kernel)  adopt temporary (if not fixed) binding approaches of threads to CPUs---with periodic migration of threads across the CPU-cores for load balancing.
Hence, having multiple Metronome threads that can become primary while managing the NIC decreases the likelihood that 
all the DPDK threads (statically pinned to different CPU-cores at startup time) share their CPU-core with higher priority interfering threads. On the other hand, we have already mentioned that Metronome---including its {\tt hr\_sleep()} architectural support---is devised with no need to explicitly impose high priority to its threads. This leaves extreme flexibility to the infrastructure owner in terms of resource-usage configuration.
The transmitting process is also influenced by batching, as DPDK moves descriptors to a transmit queue only if a certain batch threshold is reached for the same amortization reasons. This doesn't directly affect the retrieval rate, but can rather influence the latency that DPDK induces.
Transmission and receiving queues permit the host CPUs to dialog with NICs through the DMA technology: such queues usually have a variable length (on an Intel X520 NIC, users can choose a Rx/Tx queue length between 32 and 4096 descriptors).\\
In multi queue scenarios, Metronome needs to generate a random value without compromising the system performance. We leverage the DPDK's builtin Thread-safe High Performance Pseudo-random Number Generation library \texttt{rte\_random}\footnote{\url{https://www.dpdk.org/wp-content/uploads/sites/35/2019/10/Threadsafe.pdf}}.
\end{document}